\documentclass[10pt,twocolumn,preprintnumbers,amsmath,amssymb,nofootinbib
,superscriptaddress]{revtex4-1}
\usepackage{graphicx,longtable,mathrsfs,color,array}
\usepackage[hidelinks]{hyperref}
\usepackage[usenames,dvipsnames]{xcolor} 
\usepackage{amssymb,amsmath,mathtools,mathrsfs,slashed} 
\usepackage{epsfig,subfigure,placeins,float} 
\usepackage{booktabs,longtable,ctable,multirow} 
\usepackage{exscale,relsize} 
\usepackage[normalem]{ulem} 
\usepackage{enumerate}
\usepackage{enumitem}
\usepackage{comment}
\usepackage{color}
\allowdisplaybreaks[1]

\definecolor{DarkBlue}{cmyk}{1,1,0,0.1}

\begin{document}
\title{Phase effects from strong gravitational lensing of gravitational waves}
\author{Jose Mar\'ia Ezquiaga}
\email{NASA Einstein fellow; ezquiaga@uchicago.edu}
\affiliation{Kavli Institute for Cosmological Physics and Enrico Fermi Institute, The University of Chicago, Chicago, IL 60637, USA}
\author{Daniel E. Holz}
\affiliation{Kavli Institute for Cosmological Physics and Enrico Fermi Institute, The University of Chicago, Chicago, IL 60637, USA}
\affiliation{Department of Physics, The University of Chicago, Chicago, IL 60637, USA}
\affiliation{Department  of  Astronomy \& Astrophysics, The University of Chicago, Chicago, IL 60637, USA}
\author{Wayne Hu}
\email{whu@background.uchicago.edu}
\affiliation{Kavli Institute for Cosmological Physics and Enrico Fermi Institute, The University of Chicago, Chicago, IL 60637, USA}
\affiliation{Department  of  Astronomy \& Astrophysics, The University of Chicago, Chicago, IL 60637, USA}
\author{Macarena Lagos}
\email{mlagos@kicp.uchicago.edu}
\affiliation{Kavli Institute for Cosmological Physics and Enrico Fermi Institute, The University of Chicago, Chicago, IL 60637, USA}
\author{Robert M. Wald}
\email{rmwa@uchicago.edu}
\affiliation{Kavli Institute for Cosmological Physics and Enrico Fermi Institute, The University of Chicago, Chicago, IL 60637, USA}
\affiliation{Department of Physics, The University of Chicago, Chicago, IL 60637, USA}
\date{Received \today; published -- 00, 0000}

\begin{abstract}
Assessing the probability that two or more gravitational wave (GW) events are lensed images of the same source requires an understanding of the properties of the lensed images. For short enough wavelengths where wave effects can be neglected, lensed images will generically have a fixed relative phase shift that needs to be taken into account in the lensing hypothesis. For non-precessing, circular binaries dominated by quadrupole radiation these lensing phase shifts are degenerate with either a shift in the coalescence phase or a detector and inclination dependent shift in the orientation angle. This degeneracy is broken by the presence of higher  harmonic modes with $|m|\ne 2$ in the former and $|m| \ne l$ in the latter. The presence of precession or eccentricity will also break this degeneracy. This implies that a lensed GW image will not necessarily be consistent with (unlensed) predictions from general relativity (GR). Therefore, unlike the conventional scenario of electromagnetic waves, strong lensing of GWs can lead to images with a modified phase evolution that can be observed. However, we find that for a wide range of parameters, the lensed (phase modified) waveform is similar enough to an unlensed (GR) waveform that GW detection pipelines will still find it. In particular, for present detectors, we find that templates with a shifted detector-dependent orientation angle have a signal-to-noise ratio differences of less than $1\%$ for mass ratios up to 0.1, and less than $5\%$ for precession parameters up to 0.5 and eccentricities up to 0.4 at 20Hz. In these ranges, the mismatch is lower than $10\%$ with the alternative detector-independent  coalescence phase shift. Nonetheless, for a loud enough source, even with only one image it may be possible to directly identify it as a strongly-lensed image from its non-GR, phase-shifted waveform. In more extreme cases, lensing may lead to considerable distortions, and the lensed images may even be undetected with current searches. Nevertheless, an exact template with a phase shift in Fourier space can always be constructed to fit any lensed image. We conclude that an optimal strong lensing search strategy would incorporate phase information in all stages of the identification of strong-lensing, with an exact treatment in the final assessment of the  probability of multiple lensed events. This work clarifies the role that strong lensing plays in the phase evolution of GWs: how it can lead to apparent deviations from GR, how it can affect the detectability of GW events, and how it can be exploited to help identify cases of strong gravitational lensing of gravitational wave sources.
\end{abstract}

\date{\today}
\maketitle


\section{Introduction}\label{sec:introduction}
Gravitational wave (GW) astronomy was launched with the first direct detection of a binary black-hole merger \cite{Abbott:2016blz} by the LIGO/Virgo collaboration in 2015 \cite{LIGO, Virgo}. Since then, 14 events have been confirmed from observing runs O1, O2 \cite{LIGOScientific:2018mvr}, and O3 \cite{Abbott:2020uma, LIGOScientific:2020stg, Abbott:2020khf}. Now KAGRA \cite{Akutsu:2018axf} has also joined the network, and a fifth LIGO detector is expected to be built in India \cite{Aasi:2013wya}. As the sensitivities improve and the detector network expands, the number of detected events will increase dramatically, allowing for new discoveries and a deeper understanding of the universe. 

GWs are unique signals to probe the behavior of gravity in the strong field regime. As these waves propagate towards us over cosmological distance, they are sensitive to the cosmological expansion and the presence of inhomogeneities in the Universe. In particular, galaxies and clusters of galaxies can act as lenses, and lead to  multiple magnified and delayed images of GW signals. Even though the lensing probability for LIGO depends on the largely unknown merger rate of compact objects at high redshift, some scenarios predict a strong-lensing rate of 1 or more per year at design sensitivity \cite{Ng:2017yiu,Li:2018prc,Oguri:2018muv}, mostly leading to double images and about $30\%$ to quadruple images. It is important to search for possible lensed signals, as they could bias the estimation of binary and cosmological parameter. For instance, magnification would lead to an underestimation of the luminosity distance to a source, which if combined with cosmological $H_0$ constraints, would lead to an underestimation of the redshift and, as a consequence, overestimation of the source-frame chirp mass of the binary \cite{Broadhurst:2018saj}. 

One approach to identifying GW lensing is through statistical analyses over the entire population
\cite{Dai:2016igl, Broadhurst:2018saj, Oguri:2018muv, Hannuksela:2019kle, Contigiani:2020yyc, Broadhurst:2020moy}. Strong lensing would change the measured luminosity distance to the sources, and hence lead to errors in the inferred properties of black-hole mergers such as redshift and source mass distributions,
unless lensing is taken into account.

A few analyses on individual events have also been performed \cite{Broadhurst:2019ijv, Hannuksela:2019kle, McIsaac:2019use, Li:2019osa, Dai:2020tpj} (see also \cite{Pang:2020qow} for neutron stars).
These approaches assume that lensing leaves the waveform unchanged, aside from an overall magnification factor. Thus lensed events would be expected to share identical intrinsic (masses, spins) and extrinsic (sky localization, inclination) parameters, with differences present only in the inferred luminosity distances.
However, as discussed in \cite{Dai:2017huk}, lensed images come from different stationary-phase paths, and will differ by specific phase shifts in frequency space that can distort the waveform. 

In this paper, we re-analyze the interpretation and role of this lensing phase shift, and argue for the importance of including it in GW search pipelines, as well as in the assessment of strong-lensing probability for a pair of events. This phase shift was recently used for the first time as a lensing hypothesis to assess if three GW events could be images from the same source \cite{Dai:2020tpj}. In particular, \cite{Dai:2020tpj} assumed that the lensing phase shift was degenerate with a phase shift in the binary's coalescence phase, for the particular events analyzed. 
Here, we analyze in detail the conditions under which the lensing phase shift can be mimicked to good approximation by a correlated shift in the coalescence phase or orientation angle between multiple lensed images. In the simplest cases (e.g.~quasi-circular binaries with equal mass), these degeneracies maintain the shape of the waveform and
evolution of the frequency in all images, in which case lensed images are always expected to be detected with current searches. More complicated scenarios with higher modes, precession or eccentricity can lead to distortions of the waveform, especially during the late-inspiral of the coalescence. This brings up the risk of missing some of the lensed images 
if the template bank of GW searches only includes general relativity (GR) waveforms. We find that for a large sector of the parameter space of mass ratios, spins and eccentricities, 
the lensed images will not be missed as their phase evolution can be closely mimicked by GR waveforms with a value of the orientation angle that can be detector dependent (but its variation across different detectors is expected to be small in most cases). 
Extreme cases of precession may be missed by standard GR searches, but, as we will explain, even in these most general cases, a direct phase shift of the waveform in frequency space can always provide an exact description of any lensed image waveform. 
We note that when lensed events are detected, the waveform deviations from unlensed GR signals can be used to identify strongly lensed events with just a single image. Recently, this idea has been further explored for the case of 3G detectors in \cite{Wang:2021kzt}. The non-GR nature of the lensed waveforms can be used both as a way to identify lensed sources from a single image, as well as to find and confirm the strong lensing origin of multiple images.

This paper is organized as follows. In Section \ref{sec:timedelay} we start with a pedagogical description of GW lensing, where we define the stationary phase approximation for short wavelengths and introduce the lensing phase shift.
We then consider in Section \ref{sec:ToyModel} a toy model of a Gaussian wavepacket signal and illustrate the effects that a lensing phase shift will have. Readers that are familiar with lensing may skip the bulk of Sections \ref{sec:timedelay} and \ref{sec:ToyModel}.
In Section \ref{sec:Binaries} we extend previous work on lensing phase shift and include a thorough analysis of various realistic waveforms. We illustrate explicitly the differences and distortions between lensed images, and how they can be mimicked by shifts in angular parameters of unlensed waveforms. In particular, we determine the degeneracies between lensed GWs and existing GR waveform templates in binaries with asymmetric masses, spin precession and eccentricities. In Section \ref{sec:SNR} we explore the matching of these templates for different compact binary parameters, and use the signal-to-noise ratio statistic to quantify this matching. This leads us to propose in Section \ref{sec:waveform} how strongly-lensed search strategies can incorporate information on the phase to evaluate the lensing hypothesis. Finally, we discuss our results and future prospects in Section \ref{sec:discussion}.

\section{Strong Gravitational Lensing}\label{sec:timedelay}

\begin{figure}[t!]
\centering
\includegraphics[width = 0.5\textwidth]{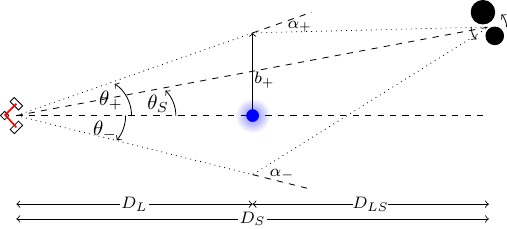}
 \caption{Diagram of strong lensing scenario, with two images. The observer (detector configuration, left) receives emission from  a GW source (compact binary, right), through a lens (blue circle, middle). Two images, with angular positions $\theta_{+}$ and $\theta_{-}$, are received by the observer. }
 \label{fig:diagram}
\end{figure}

In this section, we describe the theory of of gravitational lensing and review that, even in the regime of short wavelength, wave interference effects in strong lensing lead to a constant phase shift of the signal, which can take the values $0$, $\pi/2$ or $\pi$, depending on how the image formed. These results are valid for any kind of wave, but we emphasize its importance for GWs since the observation and parameter estimation of GWs are sensitive to the phase of the signal.

Let us consider a thin lens, i.e., a situation where lensing results from masses lying in a plane between the source and observer, such as in the point lens of  Fig.~\ref{fig:diagram}. Here, $D_L$, $D_S$ and $D_{LS}$ correspond to the comoving angular diameter distances from the observer to the lens, from the observer to the source, and from the lens to source, respectively. Also, $\vec{\theta}_S$ is the angular position of the source with respect to the lens, in the absence of the lens. Due to lensing, a source located
at an unlensed angular position $\vec{\theta}_S$ 
will have its lensed positions
$\vec{\theta}$ shifted due to all of the possible deflected paths around the lens. For a given $\vec{\theta}$, the time delay between the lensed and unlensed path in the thin lens approximation is determined by:
\begin{equation}\label{Eqtd}
    t_d(\vec{\theta},\vec{\theta}_S)\approx \frac{1}{c}\frac{D_LD_S}{2D_{LS}} |\vec{\theta}-\vec{\theta}_S|^2 + t_{ \Phi}(\vec{\theta}).
\end{equation}
The first piece is the geometric time delay and $t_{\Phi}$ is the Shapiro time delay from the time
dilation due to traversing a lensed path $s$ through the gravitational potential $\Phi$:
\begin{equation}
   t_\Phi \approx -\frac{2 }{c^3} \int \Phi ds.
\end{equation}
The resulting lensed wave at the observer position will generically be given by:
\begin{equation}
    S_L(t, \vec{\theta}_S)=\int \frac{d\omega}{2\pi} e^{-i\omega t}F(\omega,\vec{\theta}_S)S_\omega,
\end{equation}
where we have ignored the polarization of the wave, which changes negligibly due to lensing (see Appendix \ref{sec:pol}). Here, $S_\omega=\int dt e^{i \omega t} S(t)$
is the Fourier transform of the unlensed wave source $S(t)$, with time referenced to the detection epoch. 
The amplification factor \cite{Schneider:1992}
\begin{equation}\label{Feq}
    F(\omega, \vec{\theta}_s)=\frac{D_L D_S}{D_{LS}}\frac{1}{c}\frac{\omega}{2\pi i}\int d^2\theta e^{i\omega t_d (\vec{\theta},\vec{\theta}_S)},
\end{equation}
which multiplies the wave in Fourier space, takes into account all possible paths arriving to the observer. 

In the stationary phase approximation, the integral over paths is dominated by
the stationary phase points according to Fermat's principle.   
This approximation should be valid when the time delay between stationary paths is much greater than the inverse frequency of the wave \cite{NakamuraDeguchi}.
If this criterion is satisfied, then the stationary points can be viewed as corresponding to distinct images, which can be considered separately. If not, one is in a diffractive regime and the full integral in Eq.~(\ref{Feq}) would have to be computed, where wave effects can lead to distortions of the waveform \cite{NakamuraDeguchi,Takahashi:2003ix,Dai:2018enj}.
Note that for a point mass lens, the validity of the stationary phase approximation for the case where the lensing is strong enough to produce two bright images requires
that $\lambda \ll R_s$, where $\lambda$ is the wavelength of the wave and $R_s$ is the Schwarzschild radius of the point lens. 

In the case of a thin lens, the light bundles associated with the observed images can pass through at most two caustics. If there are no caustics, the image corresponds to a local minimum of $t_d$ and is said to be of type I. If there is one caustic, it corresponds to a saddle point of $t_d$ and is said to be of type II. If there are two caustics, it corresponds to a local maximum of $t_d$ and is said to be of type III.

In the stationary phase approximation, we calculate the contribution to $F$ from the $j$-th image by Taylor expanding $t_d$ as:
\begin{equation}
    t_{d}(\vec{\theta}) \approx t_d(\vec{\theta}_j)+ \frac{1}{2} \sum_{(a,b)=1}^{2}\delta \theta_a\delta \theta_b \partial_a\partial_b  t_d(\vec{\theta}_j)+ ...,
\end{equation}
where the $j$-th image is located at $\vec\theta_j$, and $\delta\vec{\theta}= (\vec\theta-\vec\theta_j)$ with two components $(1,2)$, and we
truncate the expansion of $e^{i\omega t_d}$ at the
quadratic level. This should be valid at high enough frequencies  such that
$|\omega \partial^2 t_d(\vec{\theta}_j)|^3\gg |\partial^3 t_d(\vec{\theta}_j)|^2$ (and so on for higher derivatives), 
which, in turn, should be valid when
$|\omega \Delta t_d |\gg 1$ where $\Delta t_d$ is the time delay difference between stationary points, as we previously claimed above in giving the criterion for the validity of the stationary phase approximation.
In order to obtain the resulting amplification factor in this approximation, we diagonalize the Hessian matrix $T_{ab}(\vec{\theta}_j)=(D_{LS}c)/(D_LD_S)\partial_a\partial_b  t_d (\vec{\theta}_j)$ for each image. Type I images will have 2 positive eigenvalues, type II images will have one positive and one negative, whereas type III images will have 2 negative eigenvalues. In the diagonalized basis, the amplification factor will be given by:
\begin{align}
    F\approx& \left(\frac{D_L D_S}{D_{LS}}\frac{1}{c}\frac{\omega}{2\pi i}\right)\left(\sum_j e^{i\omega t_d(\vec\theta_j)} \right)\nonumber\\
    &\times \int d^2 \tilde{\theta}  \exp\left[\frac{1}{2}i\omega \left( \tilde{\theta}_1^2 \lambda_{1j}+ \tilde{\theta}_2^2 \lambda_{2j}\right)\frac{D_LD_S}{D_{LS}c}\right],
\end{align}
where $\lambda_{1,2}$ are the two eigenvalues for each image. Renormalizing the coordinates appropriately, we can then use that $\int^{\infty}_{-\infty} dx e^{\pm ix^2}=\sqrt{\pi}e^{\pm i\pi/4}$, and obtain the final result: 
\begin{equation}\label{Fgeom}
    F\approx \sum_j\vert\mu(\vec\theta_j)\vert^{1/2}\exp\left(i\omega t_d(\vec\theta_j)-i\, \text{sign}(\omega)\frac{n_j\pi}{2}\right),
\end{equation}
where $\mu(\vec\theta_j)=\lambda_{1j}\lambda_{2j}$ is the amplification factor of the $j$-th image located at $\vec\theta_j$, and $n_j=0$ for Type I, $n_j=1$ for type II, and $n_j=2$ for type III images. $n_j$, also known as the Morse index, accounts for the phase shift that asymptotic waves far from the lens acquire due to crossing a given number of caustics between the source and the observer \cite{Schneider:1992}. 
Although, as previously mentioned, in the thin lens approximation a light bundle can cross at most two caustics \cite{Schneider:1992},
multiple lens systems could lead to additional caustic crossings, each of which would add an additional $-\pi/2$ shift. For example, crossing three caustics would lead to $-3\pi/2$, equivalent to a phase shift of $+\pi/2$ (or an overall $n^\text{tot}_j=-1$). Measuring such a phase shift would be an indication of a multiple lens system. {\em{Properties of a single lensed GW source could provide unique constraints on the nature of the lensing geometry, in a manner that cannot be done practically with electromagnetic sources.}}

Notice that the sign$(\omega)$ factor 
ensures that lensing of a real wavepacket, where 
$S_{-\omega} = S_{\omega^*}$, remains real: $F(-\omega, \vec{\theta_s}) = F^*(\omega,\vec{\theta_s})$. 
Explicitly, for a monochromatic signal $S(t)$,  the sum of  $\pm\omega$ frequency components contributes to the temporal signal of a type II  waveform $S_{\rm II}(t)$ as
\begin{align}
    S(t) &\propto  e^{-i \omega t} S_\omega + e^{+i\omega t} S_{-\omega} = 2 {\rm Re}[ e^{-i |\omega| t} S_{|\omega|} ] \nonumber \\ &= 2 |S_{|\omega|}|\cos(|\omega| t + \alpha),  \nonumber\\
    S_{\rm II}(t) & \propto 
    2 {\rm Re}[ e^{-i |\omega| (t-t_{d{\rm II}})-\pi/2} S_{|\omega|}] \nonumber \\ &= 2|S_{|\omega|}|
\cos(|\omega| (t - t_{d\rm II}) + \alpha + \pi/2) 
    \label{frequencysign}
\end{align}
where $\alpha$ is given by $S_{|\omega|}= |S_{|\omega|}| e^{i \alpha}$.  
For the type III image the temporal signal will suffer an analogous $\pi$ phase shift, which simply changes the sign of the signal.
Notice that the lensing phase shift always corresponds to a temporal phase shift of a fixed sign. 
Specifically,
a phase shift by $\alpha$ of a monochromatic wave of frequency $|\omega|$ is equivalent to shifting all its peaks and troughs in the time domain by $\Delta t = \alpha/|\omega|$. 
Thus, a real monochromatic wave is simply magnified and time shifted. Although the same phase shift applies to any frequency, a non-monochromatic waveform will be distorted in the time domain since the associated time shift of its oscillations will depend inversely on frequency, and therefore lensing will time shift each component by a different amount. As we will see later, for wave packets well localized in time, the constant lensing phase shift does not affect the group arrival time of the overall packet, but it does introduce the aforementioned time shifts of the peaks and troughs of its frequency components.

\begin{figure*}[t!]
\centering
\includegraphics[width = 0.98\textwidth]{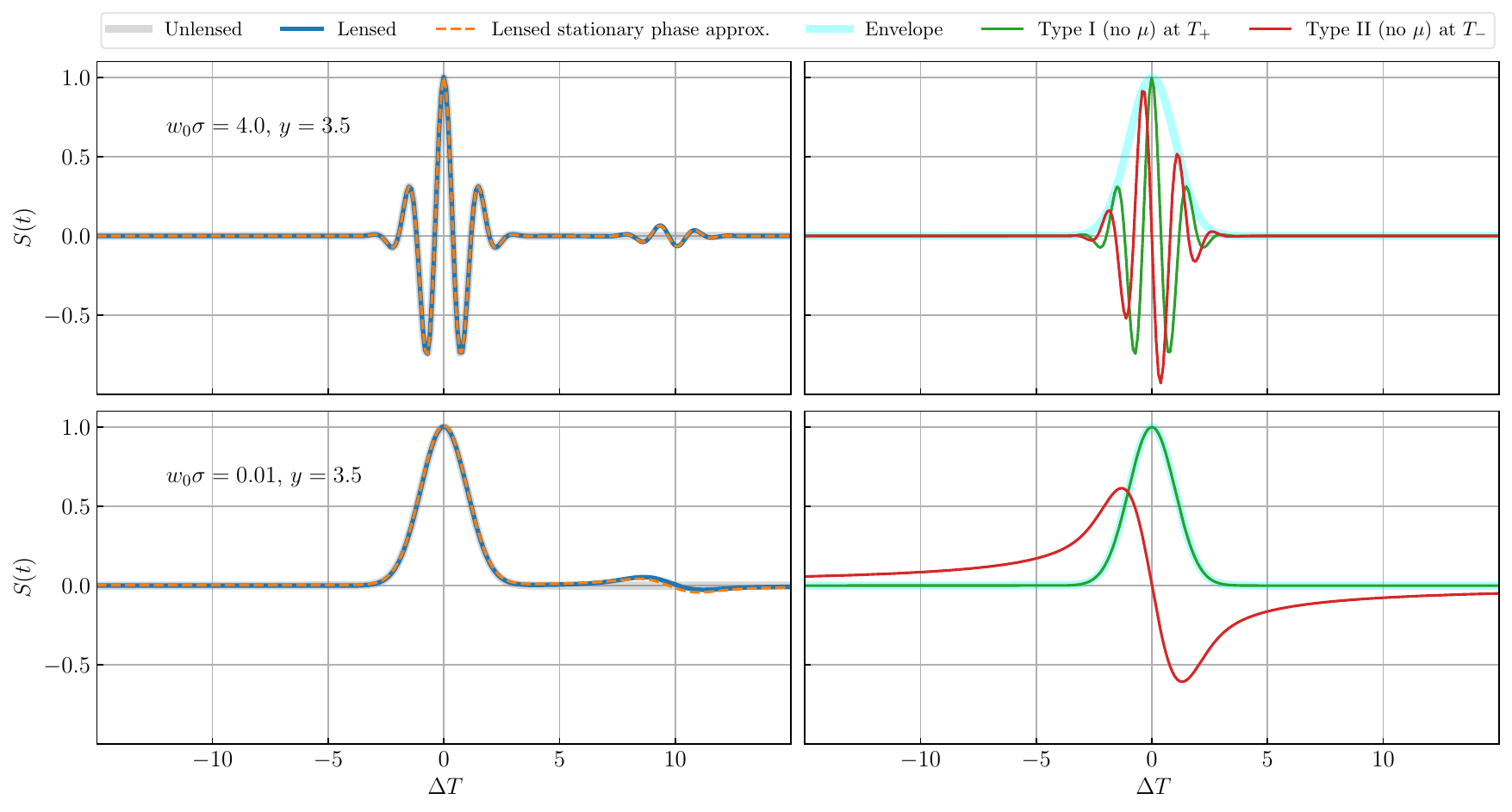}
 \caption{Effect of a point mass lens on a cosine signal modulated by a Gaussian pulse. The top panels corresponds to the high carrier frequency limit, $w_0\sigma>1$, while the lower ones represent the low frequency limit $w_0\sigma<1$. The left column presents the unlensed signal, the lensed signal computed solving the diffraction integral and using the stationary phase approximation (dashed line). On the right column we show the lensed images type I and II (without magnification) shifted to their arrival time $T_\pm$ together with the envelope of the signal. For a high  carrier frequency, we can see explicitly the phase shift introduced by lensing on the top right panel. On the other hand, in the low frequency case there is a high degree of non-monochromaticity and lensing distorts the shape of the wave-packet of the second image.}
 \label{fig:shift_cosine}
\end{figure*}

In the following we will consider the effect of the lensing phase shifts in different scenarios: first for a toy model wave-packet and later for realistic waveforms from compact binaries. For concreteness, we will focus on point lenses, but all of our conclusions hold for any thin lens for which the stationary phase approximation is valid. The only difference is that a general thin lens can generate a range of numbers and types of images, but they will all be type I, II, or III with the phase shifts previously described. Relations between the number and types of images for general lenses are given in~\cite{Blandford:1986zz,1998ApJ...495..604F}. 

\section{Lensing Shifts and Distortions}
\label{sec:ToyModel}
Now that we have reviewed the theory of strong lensing, in this section we illustrate the effect of the lensing phase shift on a pulse. We show that the lensing phase shift, since it is frequency independent, does not affect the physical arrival time of a wavepacket, and instead only determines where phase peaks and troughs of the signal are located. We also explicitly illustrate the validity of the lensing analysis performed in the previous section, showing that the time delay between multiple lensed images has to be longer than the period of the wave. The incorrect application of the lensing phase shift to such situations may lead to apparent superluminal signals that break causality.

In order to illustrate the effects of the phase shifts, we consider a wavepacket consisting of a sinusoid modulated by a Gaussian  which is  lensed by a point mass. We take the initial signal to have the form:
\begin{equation}
    S(t)= \cos{(w_0\,t/t_M)}e^{-\frac{1}{2\sigma^2}\left(\frac{t}{t_M}\right)^2},
    \label{eq:unlensed}
\end{equation}
where the dilated Schwarzschild diameter crossing time  $t_M=2R_s(1+z_L)$ defines the typical timescale of delays. We will choose the dimensionless carrier frequency in these units such that $w_0=\omega_0 t_M \gg 1$ (see
Appendix \ref{app:caustics} and \cite{Ezquiaga:2020spg} for the opposite limit). 
For a point lens, there will be two images, of type I (+) and type II ($-$), with dimensionless time delays $T_{d\pm} =t_{\pm}/t_M$
\begin{equation} \label{eq:Td}
     T_{d\pm}(y)= \frac{1}{4}\left[y^2+2\mp y\sqrt{y^2+4}\right]-\ln\,x_{\pm}
     ,
\end{equation}
where $y=\theta_S/\theta_E$ 
and 
\begin{equation}
x_{\pm}= \frac{1}{2}\left| y \pm \sqrt{y^2+4}\right|
\end{equation}
correspond to the angular source and image positions relative to the lens in Einstein ring units 
\begin{equation}
\theta_E=\sqrt{2R_s(1+z_L)\frac{D_{LS}}{D_SD_L}},
\end{equation}
and we have removed common offsets in the time delays
(see e.g.~\cite{Ezquiaga:2020spg}). 
Their corresponding magnifications are given by:
\begin{equation}
    \mu_{\pm}= \left[1-\left(\frac{1}{x_{\pm}}\right)^4\right]^{-1}.
\end{equation}

We first consider the case where $w_0 \sigma \gg 1$ so that the wavepacket is nearly monochromatic.  In Fig.~\ref{fig:shift_cosine} we take $w_0=4$, $\sigma=1$, and $y=3.5$. In the left panel, the grey line shows the unlensed image, displayed so that the maximum is  at $\Delta T=0$. The blue line shows the lensed total signal, using the expected $F$ in the stationary phase approximation in Eq.~(\ref{Fgeom}). 
The orange dotted line shows the lensed image using the full expressions for the amplification factor in Eq.~(\ref{Feq}). For this example, notice that the time delay between the two signals is large compared with the period of the carrier frequency and hence both blue and orange dotted lines coincide well.
In the left panel, the lensed signals contains both images, and this is why the signal contains two modulated Gaussians with different magnifications and time delays. As expected, the delayed image is also demagnified.

In the right panel, we highlight the phase shift of the type II image by separating the two, removing the relative magnification factor, and  aligning their arrival times in geometric optics.
For this nearly monochromatic wavepacket, we see that the type II image exhibits nearly a pure phase shift with respect to the type I image. We also show the Gaussian envelope of both images, computed as the modulus of a lensed complex signal whose real part is given by $S(t)$ (and analogous complex part with sine). We see that the envelopes coincide (i.e.~the peaks of both type I and II images touch the envelope). As explained in \cite{Ezquiaga:2020spg}, this occurs because the time delay associated to a phase shift $\Delta \Phi$ is given by $\Delta \Phi= \omega \Delta t_p$, and characterizes the shifts in arrival times of peaks and troughs of a perfectly monochromatic wave. This does not determine the physical arrival time of a temporally-localized wave packet, in analogy to phase velocity not tracking the group or front velocity of a signal. Instead, as we confirm in Fig.~\ref{fig:shift_cosine}, the group arrival time corresponding to the peak of the Gaussian envelope is given by an associated group velocity as $t_g=\partial \Phi/\partial \omega$, with the total phase given by $\Phi=\omega t_d +\Delta \Phi$. For constant $\Delta \Phi$ (as in the case of stationary phase lensing effects), the group arrival is simply given by $t_d$. The Gaussian wave packets for the type I and II images will therefore arrive at $t_{d\pm}$, 
as in the geometric optics approximation. 
Finally, notice in this example that there is also no noticeable distortion of the envelope, as the frequencies span a rather narrow range around $w_0$.  

Next we consider the opposite limit $w_0\sigma\ll 1$.
In this case, the wavepacket is nearly a pure Gaussian pulse. The lower panels of Fig. \ref{fig:shift_cosine} show the two lensed images of the Gaussian. Here we can see that the type II image (red curve) is quite distorted due to its multiple frequency components. 

For the Gaussian pulse, notice that the $-\pi/2$ phase shift of each frequency component implies that when they are re-superimposed to form the lensed wavepacket, the type II image will no longer coherently superimpose at the group arrival time $T_{d-}$ to a peak but rather to a node. Interestingly, there is a long tail to earlier arrival times. This tail results from the fact that the type II image is a saddle point of $t_d$, so parts of the wave can arrive well before $T_{d-}$.

One might worry that this tail might lead to genuine superluminality when we take the limit that the source is aligned with the lens $y\rightarrow 0$ so that $T_{d-} \approx T_{d+}$. However, in this limit, the stationary phase approximation will break down, since it requires that the time delay between images should be large compared with the period of the wave $w (T_{d+}-T_{d-}) \gg 1$. In the stationary phase approximation, the behavior of the time delay function around a saddle point is extrapolated to infinity using a quadratic expansion.  In reality, the direction to smaller delays is bounded by the global minimum that corresponds to the type I image. 
Further discussion of this falsely superluminal behavior can be found in Appendix~\ref{app:caustics}, where we conclude that its impact for a realistic gravitational wave signal is always small as long as $\omega t_M\gg 1$.

\section{Lensed Binary Inspiral}
\label{sec:Binaries}

With the understanding of the effect of lensing phase shifts in a toy model signal, in this section we proceed to consider various realistic models for GWs, and show how the lensing phase shift affects the signal in time-domain. In Section \ref{sec:geometry} we start by reviewing the standard spherical harmonic decomposition of GWs and how detectors respond to these signals. We review the physical interpretation of the main two angular parameters studied here: coalescence phase and orientation phase. In Section \ref{sec:circular} we consider the GW signal of a binary system in an idealized circular orbit. This simple model allow us to mathematically illustrate the degeneracies between the lensing phase shift and the angular parameters of waveforms. In Sections \ref{sec:circular}, \ref{sec:precession} and \ref{sec:eccentricity} we consider numerical examples of lensed waveforms in nearly circular orbits, with spin precession and eccentricity, respectively. In each section we analyze how the lensed signals can be mimicked by shifts in the coalescence and orientation phases, and why these degeneracies may break sometimes.

As we have just seen, in some cases---such as the top right panel of Fig.~\ref{fig:shift_cosine}---the waveform of a type II image may be very similar to that of the undistorted type I image. However, in other cases---such as the bottom right panel of Fig.~\ref{fig:shift_cosine}---the waveform of a type II image may be quite distorted. In this section, we consider realistic waveforms from coalescing binaries and illustrate the effect that the lensing phase shift will have. 
The key issue is to determine the degree to which the waveforms of the lensed images conform closely to unlensed waveforms---possibly with different astrophysical parameters, such as the coalescence phase and the orientation angle of the plane of the orbit. If the lensed images do not conform closely to unlensed waveforms, then they may be missed in the standard analysis or misinterpreted as confirmation of gravity effects beyond general relativity. 
If they do conform closely, then they should not be missed by the standard analysis, and the manner in which the waveforms disagree can be used to help identify the lensed images, as we shall discuss in Section \ref{sec:waveform}.

The type III waveform is simply the sign reversed type I or unlensed waveform,
and can be trivially produced in a single detector by  rotating the source polarization with respect to the detector arms by $\pi/2$
and so its detection would never be missed in the standard analysis. 
We therefore concentrate on the effects of the type II phase shift and the degree to which it exhibits degeneracies with astrophysical parameters.

\begin{figure*}[t!]
\centering
\includegraphics[width = 0.9\textwidth]{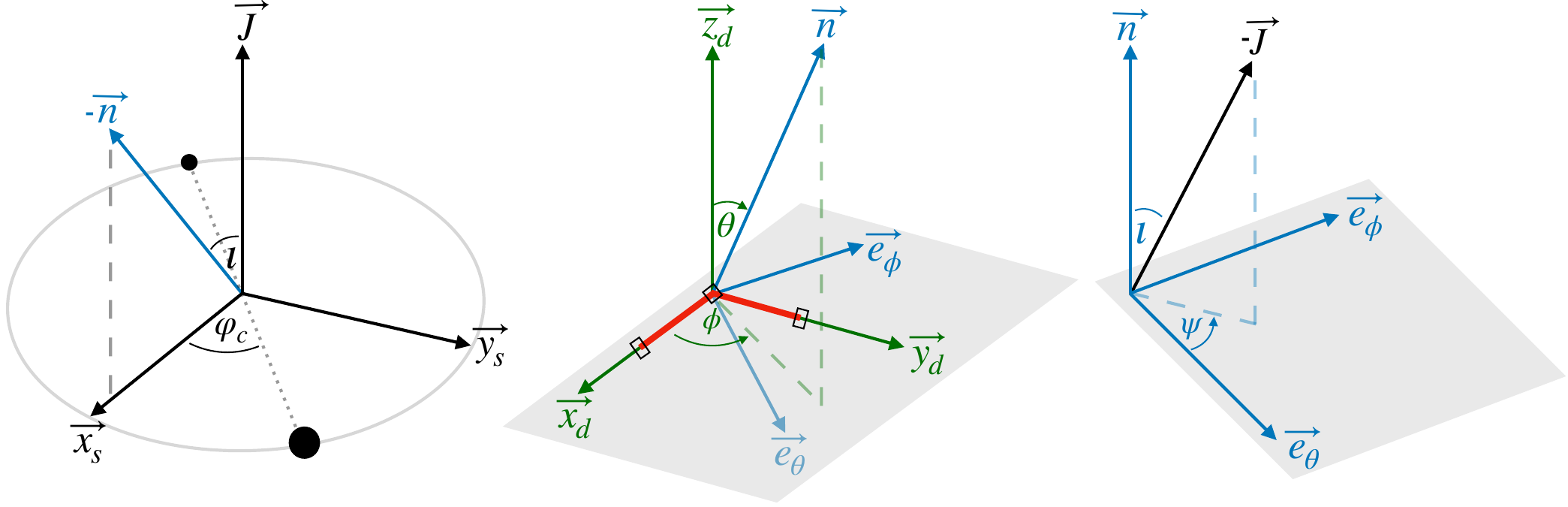}
 \caption{Diagram of source 
(black), detector (green), and sky (blue)  frames
with axes $\{\vec{x}_s,\vec{y}_s, \vec{J}\}$,  $\{\vec{x}_d,\vec{y}_d,\vec{z}_d\}$, and $\{\vec{e}_\theta,\vec{e}_\phi,\vec{n}\}$, respectively. All relevant angles are illustrated. Here, $\iota$ gives the inclination of the orbital plane with respect to the observer, where $\vec{n}$ is the line of sight direction from detector to source, and $\vec{J}$ represents the total binary's angular momentum. $\varphi_c$ corresponds to the coalescence phase. 
 In the detector frame the sky position of the source  $\{\theta,\phi\}$ are the Euler angles that align $\vec{z}_d$ with $\vec{n}$ and rotates $\{ \vec{x}_d ,\vec{y}_d\}$ to
 $\{ \vec{e}_\theta, \vec{e}_\phi \}$. $\psi$ is the final Euler angle that rotates $\vec{e}_\theta$ to the align with the transverse projection of $-\vec{J}$ and describes the orientation of the polarization. }
 \label{fig:angles}
\end{figure*}

\subsection{Source and detector geometry}\label{sec:geometry}

We choose the ``source frame'' so that $\vec{z}_s$ is oriented in the direction of the binary's angular momentum $\vec{J}$ and the observer has polar angle $\iota$ and zero azimuthal angle. We define $\varphi_c$ to be the azimuthal angle that denotes the orbital phase, i.e.~the orientation of the binary separation vector referenced to a convenient time, which we take to be coalescence. We define $-\vec{n}$ to be the direction to the observer (so that $+\vec{n}$ is the direction of the line of sight from the observer to the source). Let $+$ and $\times$ denote the two GW polarizations in the $-\vec{n}$ direction with $+$ defined by the $\vec{e}_\theta$ direction.  
We denote the amplitude of the gravitational waves in these polarizations as $h_{+}(t)$ and $h_{\times}(t)$, respectively. A general gravitational wave can be expanded as (see e.g.~\cite{RevModPhys.52.299, Kidder:2007rt}):
\begin{equation}\label{multipole}
    h_{+}(t)-i h_{\times}(t)= \sum_{\ell \geq 2}\sum_{m=-\ell}^{\ell} h_{\ell m}(t; p_i)\,{}_{-2}Y_{\ell m}(\iota,-\varphi_c),
\end{equation}
where ${}_{s}Y_{\ell m}$ denotes the spin-$s$ weighted spherical harmonic and $p_i$ denotes the parameters of the binary, which include such quantities as the masses, spins, eccentricity, and direction of the semi-major axis for eccentric binaries. We have inserted $\varphi_c$ in the argument of ${}_{-2}Y_{\ell m}$ so that $h_{\ell m}(t; p_i)$ will be independent of $\varphi_c$ for circular binaries with no spins. 
Equation (\ref{multipole}) is a completely general expansion and does not assume any properties of the source. 
All of the information about the radiation from the source is contained in the complex amplitudes $h_{\ell m}$. 

The ``detector frame'' is associated with an interferometer and is chosen so that its perpendicular arms are oriented in the $\vec{x}_d$ and $\vec{y}_d$ directions. 
Note that the angle $\iota$ introduced in the previous paragraph is also the inclination of the orbital plane with respect to the observer\footnote{In this paper, we define the inclination to be the angle between the emission direction $-\vec{n}$ and $\vec{J}$. In the convention that inclination is defined as the angle between line of sight $\vec{n}$ and $\vec{J}$, the inclination would be $\pi-\iota$.}. 
Finally, the ``sky frame'' is defined so that the $z$-axis is aligned with $\vec{n}$, and the $x$ and $y$ axes are aligned in the directions of $\vec{e}_\theta$ and $\vec{e}_\phi$, respectively. 
The relation between the detector frame and sky frame\footnote{Note that due to the earth's rotation the relation between the sky frame and the detector frame changes with time, but currently detected gravitational wave events are typically short enough that we can ignore this rotation} is determined by the
Euler angles $\{ \theta,\phi,\psi \}$ which describe
the rotation of $\vec{z}_d$ onto $\vec{n}$ and $\vec{x}_d$ onto $\vec{e}_\theta$. We employ the $zyz$ Euler angle convention where each angle represents a counterclockwise rotation around the respective axes. Note that $(\theta, \phi)$ are the polar and azimuthal angles of the source in the detector frame, whereas $\psi$, the orientation angle, describes the
orientation of $\vec{J}$ projected onto sky coordinates 
($\psi$ is sometimes also referred to as polarization angle since it affects the polarization detected).
The relation between the source
and detector frames is illustrated in Fig.~\ref{fig:angles}. A comparison of the conventions we use with other conventions in the literature is given in Appendix \ref{app:antenna}.

The strain $h$ of the detector depends on its response to the two polarizations in Eq.~(\ref{multipole}) and is given by
(see Appendix \ref{app:antenna} for a detailed derivation):
\begin{equation}
    h= F_{+}(\theta,\phi,\psi) h_{+} + F_{\times }(\theta,\phi,\psi)h_{\times},
    \label{strainobs}
\end{equation}
where $F_{+/\times}$ are the antenna pattern functions that take the explicit form:
\begin{align*}
    F_{+}&= \frac{1}{2}\left[1+\cos^2(\theta)\right]\cos (2\phi) \cos (2\psi) \\
    &-\cos(\theta)\sin (2\phi) \sin(2\psi),\\
    F_{\times} &= \frac{1}{2}\left[1+\cos^2(\theta)\right]\cos (2\phi) \sin (2\psi) \\
    &+\cos(\theta)\sin (2\phi) \cos(2\psi).
\end{align*}
Notice that under a change in $\psi$ the two linear polarization states rotate into each other 
or, equivalently, the two circular polarization states acquire a pure phase of opposite sign.

In general, the detected type II waveform will 
be a phase shifted version of Eq.~(\ref{strainobs}) which we call $h_{\rm II}$. We will see that for simple binaries, this lensing phase shift can me mimicked by a change in the orbital phase $\varphi_c$. In addition, even though the lensing phase shift does not change the polarization state, we will also see that with only one detector and for simple signals, it can be partially mimicked by a detector-dependent rotation of the polarization via $\psi$. 
In the case of type III images, it is clear from Eq.~(\ref{strainobs}) that the lensed signal will be degenerate with a $\pi/2$ shift of the orientation angle $\psi$ which induces a global sign flip of the strain, irrespective of the unlensed waveform and the detector. This means that for Type III images all the detectors will bias the orientation angle by an extra amount of $\pi/2$. In addition, since Type III images only change the overall sign of the strain and do not introduce further distortions to the waveforms, these images will never be missed with current detection analyses.

\subsection{Radiation from ideal circular motion}\label{sec:circular}

Until this point, we have made no assumptions about the source other than to set conventions on the choice of frames. Suppose, now, that we assume that the source is confined to the $z_s=0$ plane in the source frame. In that case, there will be a reflection symmetry of the gravitational radiation about this plane. It follows from this reflection symmetry that \cite{Faye:2012we}:
\begin{equation}
h_{\ell m}^* =(-1)^\ell h_{\ell -m} .
\label{reflection}
\end{equation}
On account of this relation between the gravitational radiation in $\ell,m$ and $\ell, -m$ harmonics, it is convenient to label their joint contribution by 
$\ell$, $m \geq 0$. For any such $\ell,m\ge 0$  we define $h_{+}^{\ell m}$ and $h_{\times}^{\ell m}$ by\hphantom{\,}\footnote{For $m=0$ we define $h_{+}^{\ell 0} -ih_{\times}^{\ell 0} = h_{\ell 0}\,{}_{-2}Y_{\ell 0}(\iota,-\varphi_c)$.}
\begin{equation}
\begin{split}
     h_{+}^{\ell m} -ih_{\times}^{\ell m}= h_{\ell m}\,\,&{}_{-2}Y_{\ell m}(\iota,-\varphi_c)\, +  \\
     &h_{\ell -m}\,\,{}_{-2}Y_{\ell -m}(\iota,-\varphi_c) .
\end{split}
    \label{hlmdef}
\end{equation}
The superscript $(\ell,m)$ notation denotes that 
 $ h_{A}^{\ell m}$ are not themselves multipole moments but rather the contribution of the given multipole moments to the observed signal in the emission direction.  
We refer to $h_{+,\times}^{\ell m}$ as modes, indexed by $\ell m$, e.g. $22$ for
 $\ell=2, m=2$, whereas we refer to $h_{\ell m}$ as source multipoles which carry both $\pm |m|$.  We typically call all modes beyond 22 as ``higher modes".

Now suppose that the system is in an exactly circular orbit of angular velocity $\Omega$ for all time. In that case, in the source frame, the gravitational radiation can depend upon $t$ and the source frame azimuthal coordinate $\Phi$ only in the combination $\Phi - \Omega t$. Since a multipole has angular dependence $e^{im \Phi}$ in the source frame, its value must oscillate in time as $e^{-im \Omega t}$, i.e., it will have frequency $\omega = m \Omega$. In the case of radiation emitted in the direction of a detector with angle $\Phi=0$, this means that each $\ell m$ mode will vary with $t$ and $\varphi_c$ as $e^{-im \Omega t - im \varphi_c}$. It follows from this fact together with the reflection symmetry that the modes in Eq.~(\ref{hlmdef}) take the form
\begin{align}
    h_{+}^{\ell m}&= A \cos[ m(\Omega t+\varphi_c)] ,\nonumber\\
    h_{\times}^{\ell m}&= A f_{\ell m}(\iota) \sin[m(\Omega t+\varphi_c)], \label{GWPolarizations}
\end{align}
where $A$ is an amplitude depending on inclination, distance to the source ($A\propto 1/d_L$) and the binary's properties.
Here $f_{\ell m}$, which gives the relative strengths of the two polarizations in the emission direction $\iota$, is given by
\begin{equation}
    f_{\ell m}(\iota) = \frac{{}_{-2}Y_{\ell m}(\iota,0) -(-1)^{\ell} {}_{-2}Y_{\ell-m}^*(\iota,0) }
    {{}_{-2}Y_{\ell m}(\iota,0)+ (-1)^{\ell} {}_{-2}Y_{\ell-m}^*(\iota,0) }\,,
\end{equation}
where the reflection symmetry (\ref{reflection}) has been used to obtain this formula. 
For the dominant modes of inspirals, $f_{\ell m}$ is given by 
\begin{align}
  f_{22}(\iota) &= \frac{2 \cos\iota}{1+\cos^2\iota}  ,\nonumber\\
  f_{21}(\iota) &=\cos\iota ,\nonumber\\
  f_{33}(\iota) &= f_{22}(\iota) ,\nonumber\\
  f_{32}(\iota) &= \frac{  3 \cos^3\iota -\cos\iota }{4\cos^2\iota-2} ,\nonumber\\
  f_{44}(\iota) &= f_{22}(\iota) ,\nonumber\\
  f_{43}(\iota) &= \frac{4\cos^3\iota}{6\cos^2\iota-2}.
  \label{flm}
\end{align}

The detected strain for an unlensed waveform of a circular binary therefore becomes
\begin{equation}\label{hMonochromatic}
    h= \sum_{\ell, m\ge 0} 
    \mathcal{A}_{\ell m}\cos[ m(\Omega\Delta t +\varphi_c)-\chi_{\ell m}],
\end{equation}
where $\chi_{\ell m}$ and $\mathcal{A}_{\ell m}$ are defined such that  
\begin{align} 
    &\chi_{\ell m}=\rm arctan[{F_+(\theta,\phi,\psi)},
    f_{\ell m}(\iota) {F_\times(\theta,\phi,\psi)}],\label{ChivsAngles}\\
    &\mathcal {A}_{\ell m} = A |F_{+}|[1 + \tan^2(\chi_{\ell m})]^{1/2}, \label{AvsAngles}
\end{align}
and we have used the  notation 
${\rm arctan}[x,y]= {\rm arctan}[y/x]$
 with the quadrant associated with $(x,y)$. 
Since we are describing the received signal instead of the emitted one, here we define the zero point $\Delta t=0$ to be the arrival time of the GW signal emitted at the merger time. 

A type I image will share this unlensed form 
with a change in amplitude due to magnification by a factor of $|\mu_{\rm I}|^{1/2}$ and with 
$\Delta t \rightarrow \Delta t_{\rm I}$ redefined with a zero point shifted by the time delay $t_{d {\rm I}}$:
\begin{equation}\label{hMonochromatic}
    h_{\rm I}= \sum_{\ell, m\ge 0} \vert\mu_{\rm I}\vert^{1/2}
    \mathcal{A}_{\ell m}\cos[ m(\Omega\Delta t_{\rm I} +\varphi_c)-\chi_{\ell m}].
\end{equation}
For a type II image 
\begin{align}\label{TypeIIh}
    h_{\rm II}= &\sum_{\ell, m\ge 0} \vert\mu_{\rm II}\vert^{1/2}
    \mathcal{A}_{\ell m}\nonumber\\
    &\times\cos\left[ m\left(\Omega \Delta t_{\rm II}+\varphi_c\right)-\chi_{\ell m}+\frac{\pi}{2}\right],
\end{align}
where again $\Delta t_{\rm II}$ is shifted by
the appropriate time delay $t_{d{\rm II}}$ and the amplitude rescaled by the appropriate magnification factor. The term $\pi/2$ in the argument of the cosine factor is the 
lensing phase shift given by Eq.~(\ref{frequencysign}). 

By comparing Eq.~(\ref{hMonochromatic}) to Eq.~(\ref{TypeIIh}), we see that we can partially mimic the effect of the lensing phase shift of $\pi/2$ by either a shift in the coalescence phase, $\varphi_c$, or by a change in the orientation angle $\psi$ (which affects $\chi_{\ell m}$), or both. We analyze the effect of each one of these angles separately. Let us first consider the effect of a shift in $\varphi_c$. If only one $m$-value were to contribute to $h_{\rm II}$, then a phase shift of $\Delta \varphi_c=\pi/(2m)$ 
in Eq.~(\ref{hMonochromatic}) would exactly mimic the lensing phase shift of the type II image. In the case of equal-mass binaries, where the radiation is dominated by $m=2$ modes, the phase of the type II image can be obtained to a good approximation
simply by shifting $\Delta \varphi_c=\pi/4$. However, for binaries with unequal masses, higher $m$ emission will be important, so a shift in $\varphi_c$ will not, in general, reproduce the constant phase of the type II lensed image in Eq.~(\ref{TypeIIh})\footnote{Much more generally, a coalescence phase shift of $\pi/4$---together with a $\pi/4$ rotation of any other binary parameters that depend on direction for the case of non-circular or spinning binaries---will mimic the type II lensing phase shift whenever (i) the sum in Eq.~(\ref{multipole}) is dominated by $m = \pm 2$ and (ii) the time Fourier transform of $h_{\ell 2} (t)$ has only positive frequencies whereas the time Fourier transform of $h_{\ell -2} (t)$ has only negative frequencies. (Similarly, a coalescence phase shift of $-\pi/4$ will mimic the type II lensing phase shift when these conditions hold with positive and negative frequencies interchanged.) For circular motion, only the frequencies $\omega = m \Omega$ are present, so condition (ii) holds automatically for $\Omega > 0$.}. 
Nevertheless, in the special case of face-on $\iota=0$ $(\pi)$ binaries in circular motion 
the GW emission will be purely circularly polarized, i.e.,
only $m=2$ ($m=-2$) harmonics 
contribute in Eq.~(\ref{multipole})  for any $\ell$. In this case, the coalescence phase shift $\Delta \varphi_c=\pi/4$ will exactly reproduce the lensing phase shift. 

We now consider changing the orientation angle $\psi$, which has the effect of rotating the plane of the orbit around the line of sight. 
From Eq.~(\ref{ChivsAngles}), we see that a shift in $\psi$ will induce a shift in $\chi_{\ell m}$. If $\Delta \psi$ can be chosen so that $\Delta \chi_{\ell m}=-\pi/2$, we will reproduce the phase of a type II lensed image. Notice, however, that a shift in $\psi$ will also rescale the amplitude of the signal according to Eq.~(\ref{AvsAngles}), which can be interpreted as a change in the luminosity distance. 

We emphasize that, for type II images, both the change in $\psi$ and amplitude will depend on angular parameters relative to the detector plane, and therefore different detectors may relate the two lensed images by different shifts of parameters. Since shifts in $\psi$ change the polarization states of the wave, then if multiple detectors are able to constrain independently both polarizations (and hence constrain $\psi$ well enough), then the detected type II signal will no longer be degenerate with a lensing phase shift. In Appendix \ref{app:antenna} we discuss the extent to which 
the type II image will be degenerate with a orientation shift $\Delta\psi$ for multiple detectors, which depends mostly on the binary inclination and works well in general except for binaries close to edge on. In the rest of the paper, we focus on the degeneracies for a single detector scenario. Note, however, that even for a single detector, the time delay between the multiple images will cause the detector to change its location and orientation with respect to the binary source. Therefore, formally, the phase shift degeneracies discussed here will hold with respect to an inertial frame that does not rotate with Earth. In principle, one would have to compute $\Delta\psi$ with respect to the source-detector geometry at the time of arrival of each image (which can be achieved using the detectors coordinates and GPS time). In practice, the parameter estimation from each event will be typically quoted with respect to the same inertial frame, where both events localizations should agree, and $\Delta\psi$ can be calculated using those localization constraints, and the $\psi$ constraint of the earliest image.

In the equal-mass binary case, the radiation is dominated by the 22 mode so that there is a single change in $\psi$ that mimics lensing. Explicitly, we have
\begin{widetext}
\begin{equation}\label{DeltaPsim2}
    \tan(2\Delta\psi_{22})=\frac{\cot(\chi)+f_{22}(\iota)\tan(2\psi)+f_{22}(\theta)\tan(2\phi)(f_{22}(\iota)-\cot(\chi)\tan(2\psi))}{\cot(\chi)(f_{22}(\theta)\tan(2\phi)+\tan(2\psi))+f_{22}(\iota)(f_{22}(\theta)\tan(2\phi)\tan(2\psi)-1)}\,.
\end{equation}
\end{widetext}
To obtain $\Delta \psi_{22}$ itself, we employ the arctan
branch solution where
$\Delta\psi_{22} \in (-\pi/2,0)$
if $\cos(\iota)>0$ else $\in (0,\pi/2)$. 
In addition, the distance with respect to the type I image will be rescaled as
\begin{equation}
    d_L \to \frac{\mathcal{A}(\theta,\phi,\psi+\Delta\psi_{22},\iota)}{\mathcal A(\theta,\phi,\psi,\iota)}\,d_L\,\vert(\mu_{-}/\mu_{+})\vert^{-1/2}.
\end{equation}

If higher modes contribute significantly to the radiation, a common $\Delta \psi$ shift will still mimic the type II lensing phase shift if the dominant modes present have the same $f_{\ell m}$. 
From Eq.~(\ref{flm}) we can see that if the dominant modes are such that  $m=\ell$,
then there will be a degeneracy between a shift in $\psi$ and lensing. Thus, a shift in $\psi$ can be expected to provide a better approximation than a shift in $\varphi_c$ to the the type II lensing phase shift in the case where $\ell > 2$ modes with $m=\ell$ are significantly present.

In the case of a face-on binary, $\psi$ and $\varphi_c$ are degenerate, i.e., a rotation along the line of sight is equivalent to a rotation in the orbital plane.
Equivalently, since the emission is a pure circular polarization state, it acquires a pure phase under rotation like the lensing effect.
Thus, in the face-on case, a rotation by $\Delta \psi = \mp\pi/4$ will exactly reproduce the lensing phase shift (where the sign depends on $\iota=0$ or $\pi$ respectively). 
\footnote{It is to be noted that in previous work \cite{Dai:2017huk} the degeneracy of the rotated 22 mode with a type II image was phrased in terms of an ``azimuthal angle" $\psi$ that corresponds to our coalescence phase $\varphi_c$ - L. Dai private communication.}

Finally, we also could allow $\varphi_c$ and $\psi$ to vary simultaneously. A degeneracy with the type II image will occur if $\Delta\varphi_c$ and $\Delta\psi$ are such that $\Delta \chi_{\ell m}=-\pi/2 + m\,\Delta\varphi_c$.

\begin{figure*}[t!]
\centering
\includegraphics[width = 0.98\textwidth]{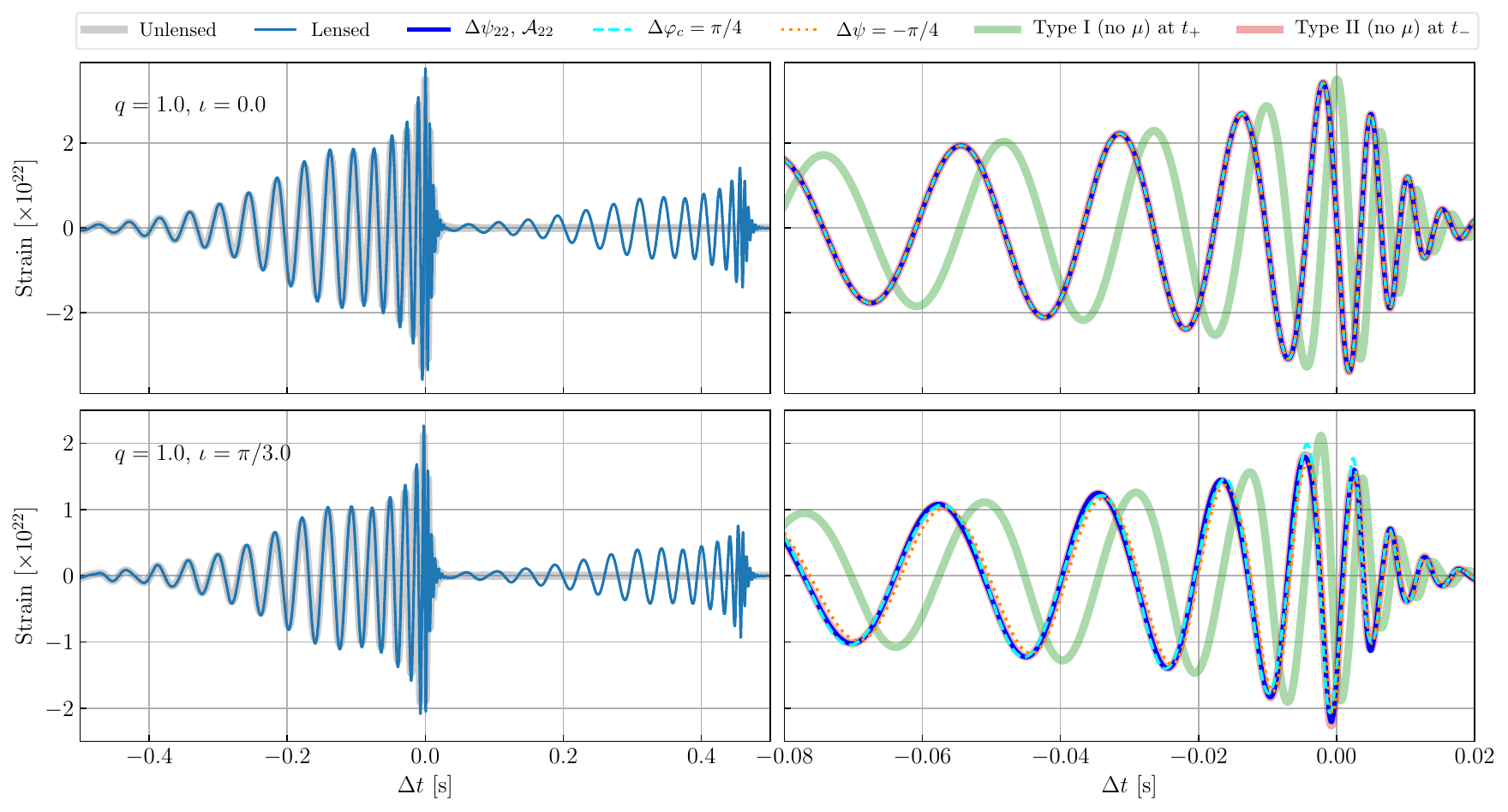}
 \caption{Effect of a point lens on a GW signal from a non-spinning, circular binary with equal masses ($q=1$). The upper panels correspond to a face-on binary with inclination $\iota=0$ while the lower ones have $\iota=\pi/3$. The left panels present the unlensed and lensed signals. The right panels show the type I and II images (without magnification and shifted to remove their time delays) together with the unlensed signal of the same binary but with an orientation and amplitude change according to Eq.~(\ref{DeltaPsim2}) (solid blue line). We also show the unlensed signal with a shift in the coalescence phase (dashed line) but with the true orientation (dotted line). 
In all the panels we have chosen $\mathcal{M}_c=26.1 M_\odot$, $z_S=0.5$, $M_L=10^4M_\odot$, $z_L=0.1$, $\theta=0.3$, $\phi=0.4$ and $\psi=1.5$. The unlensed waveforms have been computed using \texttt{IMRPhenomHM} \cite{PhysRevLett.120.161102}, which contains higher modes.}
 \label{fig:wave-formq1}
\end{figure*}

\subsection{Nearly circular inspiral}
\label{sec:circular}

The analysis of the previous subsection shows that for ideal circular motion, if the signal is dominated by $m=2$ modes, the lensing phase shift of a type II image can be mimicked by a coalescence phase shift of $\Delta \varphi_c=\pi/4$. Alternatively, if the signal is dominated by $m=\ell$ modes, the lensing phase shift can be mimicked by the rotation Eq.~(\ref{DeltaPsim2}) about the line of sight. We now turn to the consideration of how well the type II image is mimicked by these changes in $\varphi_c$ and $\psi$ for realistic binary inspirals. In this subsection, we consider nearly circular inspirals with no spin, where the previous circular results still apply for a time-varying binary's angular velocity $\Omega(t)$. In the next subsection, we consider binaries with spin (which causes the orbital plane to precess) and eccentric orbits.

\begin{figure*}[t!]
\centering
\includegraphics[width = 0.98\textwidth]{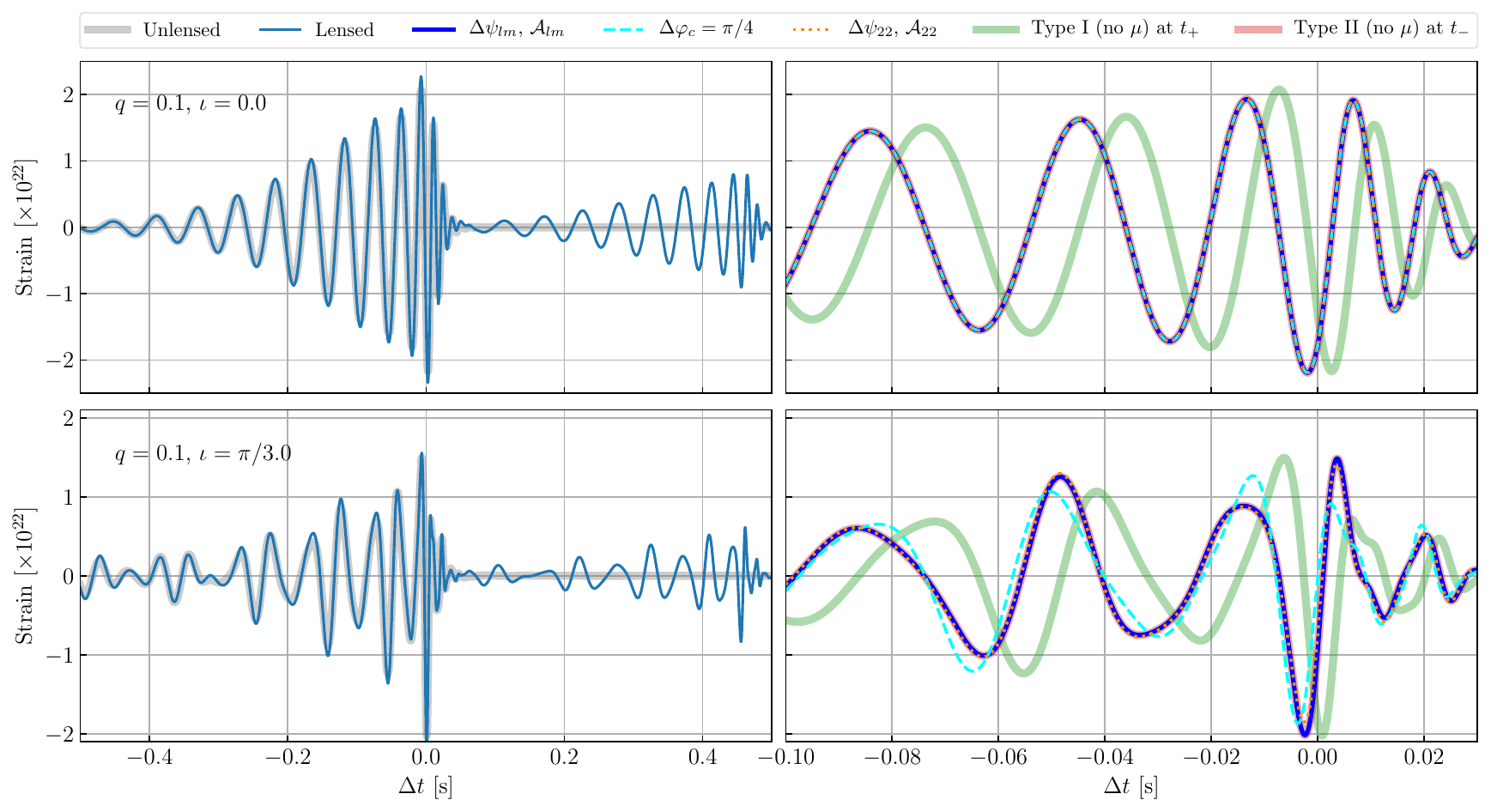}
 \caption{Effect of a point lens on a GW signal from a non-spinning, circular binary with unequal masses ($q=0.1$). The upper panels correspond to a face-on binary with inclination $\iota=0$ while the lower ones have $\iota=\pi/3$. We plot the same $\Delta\varphi_c$ and $\Delta\psi_{22}$ lines as in Fig. \ref{fig:wave-formq1}. In addition, we include the unlensed signal where the orientation and amplitude have been changed for each $(\ell,m)$ to mimic the phase shift of a type II image (see Eq.~(\ref{ChivsAngles})). The rest of the parameters are the same as in Fig. \ref{fig:wave-formq1}. The unlensed waveforms have been computed using \texttt{IMRPhenomHM} \cite{PhysRevLett.120.161102}, which contains higher modes.}
 \label{fig:wave-formq008}
\end{figure*}

Fig.~\ref{fig:wave-formq1} shows the GW signal of a binary with equal-mass non-spinning black holes in a circular orbit, when lensed by a point mass $M_L=10^{4}M_\odot$ located at $z=0.1$. The source is located at $\theta =0.3$ and $\phi=0.4$ radians.
Following the same plotting conventions as the previous figures, on the left panels of Fig.~\ref{fig:wave-formq1} we show the unlensed and lensed total GW signal, where we have aligned $\Delta t=0$ and $\Delta t_{\rm I}= 0$ in order to compare their profiles straightforwardly. On the right panels we show separately the two lensed images type I and II, aligning $\Delta t_{\rm II}=0$ as well. There, we have ignored the relative magnifications of each image. 
Moreover, in order to emulate the strain detected by present ground-based detectors, we set a lower frequency cutoff at 25Hz. An upper cutoff is not necessary since for the choice of chirp mass the binary merges in band. We apply this filter to both the unlensed and lensed signals and this is why in the left panels each image has a duration of $\sim0.5$s.

The top panel shows the case of a face-on binary, $\iota=0$, and the bottom panel shows a binary with $\iota=\pi/3$. In both cases we see that the type II image has a phase shift compared to the type I, and no distortions in the waveform are visible. Because the GW signal during the inspiral is dominated by $m=2$ modes, and in particular $\ell=2$, the type-II image is well matched to the type-I with a shift in the coalescence phase parameter by $\pi/4$
in the strain. In the case of $\iota=\pi/3$, during the late inspiral there will be other modes contributing, such as $\ell=m=4$ and thus for the signal shifted by $\Delta \varphi_c=\pi/4$ the match is not perfect. We will quantify these differences more quantitatively in terms of the signal-to-noise ratio in the next section.

Alternately, one can instead shift the orientation angle together with an amplitude rescaling, depending on the inclination of the source and location. For a face-on binary ($\iota=0$), there is no need to rescale the amplitude, and a shift in the orientation angle will be equivalent to a shift in the coalescing phase regardless of the location of the source, as previously explained. This is confirmed in the top panels of Fig.~\ref{fig:wave-formq1}.
For the case of finite inclination, we can use Eq.~(\ref{DeltaPsim2}) to obtain the shift in $\psi$ needed to reproduce the lensed type-II image. In the bottom panel of Fig.~\ref{fig:wave-formq1}, we first confirm that the shift $\Delta\psi=-\pi/4$ does not match perfectly the phase evolution of the type II signal.  
Instead, if we perform a shift by $\Delta\psi_{22}=-0.28\pi$ and a fractional change in amplitude by $\Delta\mathcal{A}_{22}/\mathcal{A}=1.1$, the signal matches well the lensed type II image during the inspiral as predicted.   

\begin{figure*}[t!]
\centering
\includegraphics[width = 0.98\textwidth]{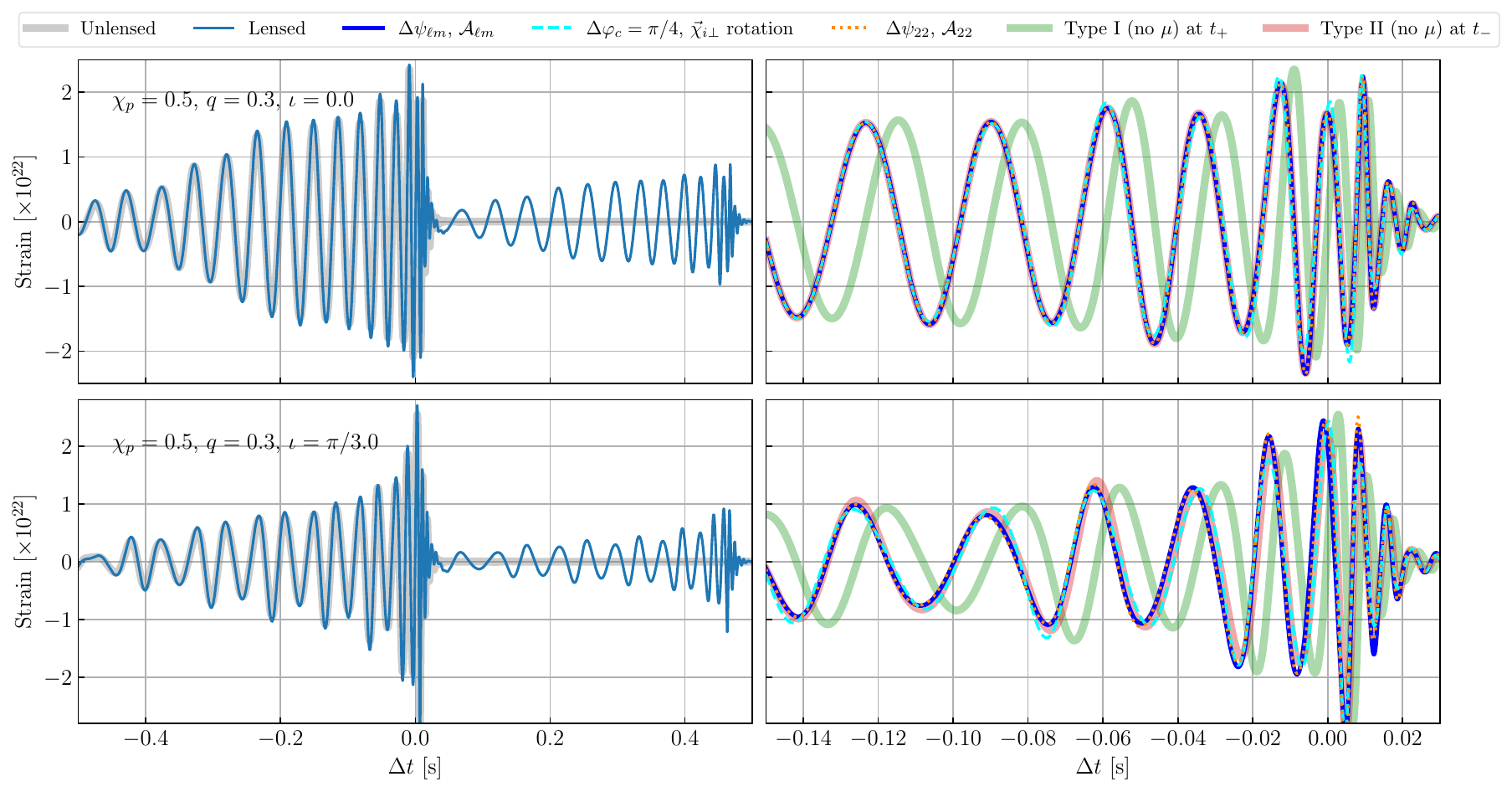}
 \caption{Effect of a point lens on a GW signal from a spinning ($\chi_\text{eff}=0.5$), circular binary with unequal masses ($q=0.3$) and exhibiting precession ($\chi_p=0.5$). The upper panels correspond to a face-on binary with inclination $\iota=0$ while the lower ones have $\iota=\pi/3$. We use a phenomenological waveform model \texttt{IMRPhenomPv3HM} \cite{Khan:2019kot} which contains higher modes in the co-precessing frame. We plot the same lines as in Fig. \ref{fig:wave-formq008} but when we shift the coalescence phase by $\Delta\varphi_c$ we also rotate the perpendicular projection of the spins $\vec{\chi}_{i\perp}$ by $-\Delta\varphi_c$. The rest of the parameters are the same as in Fig. \ref{fig:wave-formq1}. }
 \label{fig:wave-form_precession}
\end{figure*}

Next, we consider the case of a circular planar binary with highly asymmetric masses. 
Fig.~\ref{fig:wave-formq008} shows the GW signal in the case of a mass ratio of $q=0.1$, with the same angular and lens parameters as Fig.~\ref{fig:wave-formq1}. The top panels show the case of a face-on binary. In this case the type II image exhibits no relevant distortion of the waveform with respect to the type I one, as the shape of the envelope of the GW signal is unchanged. 
As previously discussed, this occurs because the signal contains only $m=2$ modes by symmetry, and therefore, during the inspiral, the strain contains a single temporal oscillating component with a slowly-varying frequency $2\Omega$ so that lensing simply shifts the waveform.  
Here, we confirm that a constant shift by $\Delta \psi=-\pi/4$ 
or $\Delta \varphi_c=\pi/4$ is degenerate with the lensed image type II. 

The bottom panels of Fig.~\ref{fig:wave-formq008} show the lensed waveform when the inclination is $\iota=\pi/3$. In this case, higher $m$ modes will have a significant contribution to the strain, and thus we expect a constant phase shift to distort the signal. 
A shift in $\varphi_c$ will therefore not reproduce the correct signal of the type II image. This result is confirmed by the cyan curve in the right bottom panel of Fig.~\ref{fig:wave-formq008}.

If we instead make a shift $\Delta \psi_{22}$ together with a rescaling $\Delta \mathcal{A}_{22}$ with the values of Fig. \ref{fig:wave-formq1} (we are using the same source position, just with $\iota=\pi/3$), we find that the new signal matches the type II image very well (shown in dotted line). This is because the modes $m=\ell$ are the dominant ones 
and when only those modes are present, there is a single shift $\Delta \psi$ and rescaling $\Delta \mathcal{A}$ that is can mimic the type II image. We will see in Sec.~\ref{sec:SNR} that the difference in the template matching is less than $1\%$. Since the signals are never 100$\%$ dominated by $m=\ell$ modes (even in extremely symmetric cases like equal masses), then a better match to the type II image can be obtained by making mode-dependent shifts $\Delta \psi_{\ell m}$ and corresponding rescalings $\Delta \mathcal{A}_{\ell m}$ for each one of the modes present. In the bottom panels, we confirm that this matches very accurately the type II image (see dark blue line). 
For reference, the orientation shift of the other modes present is $\Delta \psi_{21}=-0.32\pi$, $\Delta \psi_{32}=-0.35\pi$ and $\Delta \psi_{43}=-0.25\pi$ (to be compared with $\Delta \psi_{22}=-0.28\pi$ for $m=\ell$ modes).

In summary for nearly circular inspiral, the type II lensed image will be well matched either by a change in $\varphi_c$ (if $m=2$ modes dominate) or by a change in $\psi$ and amplitude (if $m=\ell$ modes dominate). 
In the fully general case, a change in $\psi$ and amplitude per mode would provide an almost perfect match for each detector. Note that the match is never exactly true since the binary is going through inspiral, merger and ring-down, so that Eq.~(\ref{hMonochromatic}) does not hold all the way. Nonetheless, we have seen that the degeneracy for practical considerations is essentially exact.

\subsection{Precession}
\label{sec:precession}

For scenarios where the spins $\vec{S}_i$ of the masses are relevant, there could be precession in the orbit and the GW emission resulting in further departures from the behavior, Eq.~(\ref{GWPolarizations}), of an ideal circular binary. Nevertheless, for nearly circular binaries, the GW emission in the non-inertial co-precessing frame defined by the orbital angular momentum $\vec{L}$, the mode emission is still well-approximated by that of a non-precessing binary \cite{PhysRevD.84.024046, PhysRevD.88.024040}. This frame precesses around the inertial frame, defined by the total angular momentum $\vec{J}$, 
and should have an effect similar to making the 
effective viewing angle, $\iota$,  and polarization orientation $\psi$ change with the precession.
 Therefore, the GW mode emission in the source frame can be well approximated by time dependent
rotation of the emission direction in the co-precessing frame, instead of the fixed direction $\iota$ and orientation $\psi$ of the non-precessing case \cite{Hannam:2013oca, Khan:2019kot}. 
Therefore, it should be expected that the degeneracies of the lensing phase shift with $\Delta \psi$ of the circular planar case, should be well described by time dependent ones that track the rotation of the emission angle.  In fact, we shall find below that that errors induced by using the fixed, circular planar analysis remain small unless precession rotates the orbit to approach or cross the edge-on limit.

In Fig.~\ref{fig:wave-form_precession}, we use  the phenomenological waveform \texttt{IMRPhenomPv3HM} to simulate realistic waveforms for precessing binaries \cite{Khan:2019kot} 
(see Appendix \ref{app:waveform} for further details). 
For circular binaries, the best measured combination of spin parameters is the effective dimensionless spin
\begin{equation}
    \chi_\text{eff}=\frac{\chi_{1\parallel} + q\cdot\chi_{2\parallel}}{1+q}\,,
\end{equation}
which is a mass weighted projection into the orbital angular momentum ($\chi_{i\parallel}=\vec L\cdot\vec\chi_i$) where we have defined the dimensionless spins $\vec\chi_{1,2} = \vec S_{1,2}/m_{1,2}^2$. On the other hand, the amount of precession can be parametrized with the precessing spin parameter \cite{Schmidt:2012rh}
\begin{equation} \label{eq:chi_p}
    \chi_p = \text{max}\left(\chi_{1\perp},\frac{4q+3}{4+3q}q\,\chi_{2\perp}\right)\,,
\end{equation}
where $\vec\chi_{i\perp}=\vec\chi_i-\vec\chi_{i\parallel}$. This parameter quantifies the amount of spin in the plane perpendicular to $\vec L$. For this figure we will consider the simple choice 
in which $\chi_{1\perp}=\chi_p$, $\chi_{1\parallel}=0$, $\chi_{2\perp}=0$ and $\chi_{2\parallel}=\chi_\text{eff}$. 

 Fig.~\ref{fig:wave-form_precession} shows the lensed waveform of a precessing GW signal with $\chi_p=0.5$ and mass ratio $q=0.3$, for the same lens parameters as the previous figures.
Typically, the precession timescale is longer than the orbital timescale, in which case precession will mainly induce a modulation of the amplitude, which can be seen on the left plots.

In the top panels, we show the case of a binary with $\iota=0$ (i.e.~with $-\vec{n}$ and $\vec{J}$ aligned). 
Even for an $\iota=0,\pi$ inclination of $\vec{J}$, 
the polar viewing angle relative to $\vec{L}$ no longer vanishes and thus the emission 
can contain $m \neq 2$ modes. 
Therefore, a shift in the coalescence phase, even in this case, is only approximately equivalent to the lensing phase shift and depends on the amplitude of the higher modes, which is more substantial 
here given the unequal masses. 
On the right plot we confirm this expectation by considering the effect of a constant shift in $\Delta \varphi_c=\pi/4$ together with a $-\pi/4$ rotation of the perpendicular dimensionless spins\footnote{The rotation of $\vec{\chi}_{i\perp}$ is needed because
the spin directions are conventionally defined relative to the position of the masses at a fiducial epoch. Note that this rotation of spins does not change the values of $\chi_p$ and $\chi_\text{eff}$ and, therefore, both images will be characterized by the same physical spin parameters.} $\vec{\chi}_{i\perp}$. We see that these shifts give 
a good match of the type-II image during the early inspiral when the (22)-mode completely dominates. The small differences observed are indeed caused by the presence of higher harmonic modes, as we have confirmed that if we only include the (22)-mode of the co-precessing frame (for example using the approximant \texttt{IMRPhenomPv3} \cite{Khan:2018fmp}) the match between the $\Delta \varphi_c$ shift and the lensed type-II signal is nearly exact.
These differences also appear in the $\iota=\pi/3$ case shown in the bottom panel where the effect of the higher modes are even more visible (and hence the matching of the type II image signal with $\Delta\varphi_c = \pi/4$ is worse), as was the case for non-precessing binaries with larger inclinations.

\begin{figure*}[t!]
\centering
\includegraphics[width = 0.98\textwidth]{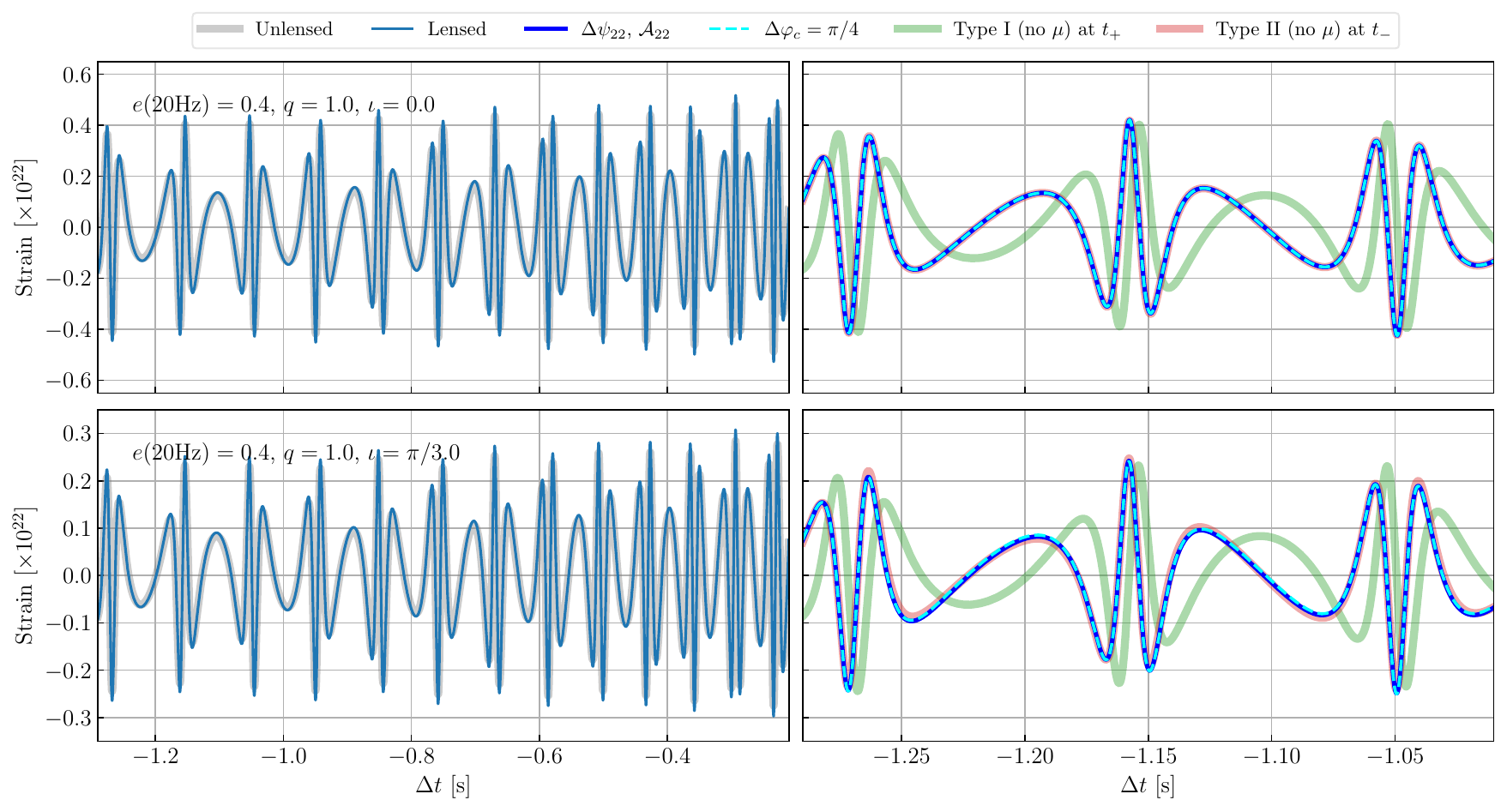}
 \caption{Effect of a point lens on the GW inspiral from a non-spinning, eccentric, equal masses binary with $e=0.4$ at 20Hz. On the left we plot the inspiral of the first image compared to the unlensed signal, while in the right we present the different images (rescaled to their arrival time) in a snapshot of the early inspiral. The upper panels correspond to a face-on binary with inclination $\iota=0$ while the lower ones are inclined with $\iota=\pi/3$. We use the waveform model \texttt{EccentricTD} \cite{Huerta:2014eca} describing the quadrupolar radiation  
 during the inspiral phase only, up to the inner most stable circular orbit. We plot the same lines as in Fig.~\ref{fig:wave-formq1}, using the same lens parameters (except $M_L=10
^{5.5}M_\odot$) with a chirp mass of $8.7M_\odot$. 
}
 \label{fig:wave-form_eccentricity}
\end{figure*}

The lensing degeneracy with $\psi$ also changes in two ways. The first is that the ratio of polarization states characterized by $f_{\ell m}(\iota)$ should be generalized to reflect the viewing angle 
with respect to $\vec{L}$, thus changing the best value to $\Delta\psi\ne \Delta\psi_{22}$. The second is that this angle varies slowly over the precession cycle.  If the viewing angle passes close to an edge-on orbit configuration then the degenerate 
$\Delta\psi$ can change rapidly. 

For the $\iota=0$ case, 
the effective viewing angle would only approach orbital edge on for extreme precession angles (e.g.~when $S_{1,2}\geq L$).  
Therefore, in this example,
the non-precessing predictions 
$\Delta\psi_{22}$ and ${\cal A}_{22}$
still match the type II lensing image quite well. 

For the inclined $\vec{J}$ case $\iota=\pi/3$, the precession of $\vec{L}$ rotates the effective viewing angle to being closer to edge on. There we see that $\Delta\psi_{22}$ and ${\cal A}_{22}$ no longer perform as well. 
For the same reason, the predicted shifts of each mode $\Delta\psi_{\ell m}$ from the non-precessing case no longer produce a nearly exact degeneracy.  
Note that in these examples, the shift in $\Delta \psi_{22}$ and $\Delta \psi_{\ell m}$ reproduce almost the same waveform, indicating that the main effect causing the deviation from the type-II image is the error in calculating the appropriate $\Delta \psi$ only, instead of the presence of higher modes.

One should note that different mappings between $\vec{\chi_1},\,\vec{\chi_2}\to\chi_\text{eff},\,\chi_p$ could give different levels of match for each curve, but the qualitative behavior is the same.
In Sec.~\ref{sec:SNR},  we will quantify this more precisely by computing the matched filter signal-to-noise ratios of different GW signals.

\subsection{Eccentricity}
\label{sec:eccentricity}

Here we consider the effect of eccentricity. Eccentric binaries are (nearly) periodic, so the GW radiation must still occur at integer multiples of the orbital frequency $\Omega$. However, the radiation need not be dominated by $\omega = m \Omega$
(see e.g.~\cite{Martel:1999tm, Hinder:2008kv, Huerta:2016rwp}). In particular, the dominant components to the strain during early inspiral will be at frequencies $\Omega$, $2 \Omega$, and $3 \Omega$ (and they are all present even for the 22 modes), with amplitudes that depend on the eccentricity $e$ and can enhance the overall strain amplitude and SNR \cite{Yunes:2009yz}. In addition, the fact that $\Omega$ is not uniform and effects such as pericenter precession introduce modifications in the binary's evolution and hence the GW phase, when compared to a circular binary. During coalescence, the orbit circularizes and therefore eccentricity is expected to have larger observable effects during the early inspiral.

For simplicity, let us focus only on $\ell=2$ modes. The aforementioned modulations in the amplitude and phase evolution of the strain can be seen on the left panels of Fig.~\ref{fig:wave-form_eccentricity}, where we show an equal-mass binary system with initial eccentricity $e=0.4$ at 20Hz, for two different inclinations $\iota=0$ and $\iota=\pi/3$. We only present the inspiral phase of the first image, since we use the numerical \texttt{EccentricTD} waveforms \cite{Huerta:2014eca} that do not describe merger and ring-down. This waveform model cuts the GW signal at the inner most stable circular orbit (ISCO), $f_\text{isco}\approx 146.6 (30M_\odot/(1+z)M_\text{tot})$Hz. In this example, the second image will have a time delay $\Delta t_d=14$ sec. 

For face-on binaries, the 22 modes dominate. In the case of circular binaries viewed face-on, a shift of $\varphi_c$ is equivalent to a shift of $\psi$. However, for eccentric binaries viewed face on, a shift in $\varphi_c$ changes the location of the masses with respect to the semi-major axis---thereby changing the physical system---whereas a shift in $\psi$ does not. Therefore, for face-on eccentric binaries, we would expect a shift in $\psi$ to perform better than a shift in $\varphi_c$ with regard to mimicking a type II image. Indeed, we find that a shift of $\Delta \psi = -\pi/4$ does indeed reproduce the type-II lensed image, as shown in the top right panel of Fig.~\ref{fig:wave-form_eccentricity}. Although it is not visible by eye, we have checked that the $\Delta\varphi_c = \pi/4$ shift does not match exactly while $\Delta \psi = - \pi/4$ shift does provide an almost perfect match. 

In the case of inclined binaries, additional terms with frequency $\Omega$ and different ratio $h_{+}/h_{\times}$ will contribute to the strain (more specifically, these terms come from the $\{\ell=2,m=0\}$ mode), and will cause the degeneracy with $\Delta \psi_{22}$ and the type-II image to also break during the early inspiral. This is confirmed in the bottom right panel of Fig.~\ref{fig:wave-form_eccentricity}, where we see that for an inclination of $\iota=\pi/3$ the match is worse than in the top panels for both cases.

It is interesting to note that for this range of eccentricity parameters the approximate $\Delta\varphi_c$ and $\Delta\psi_{22}$ degeneracies, although not perfect, still mimic well the lensed type II image. 
If $\ell>2$ modes were included in the waveform approximant, additional deviations with respect to the $\Delta\varphi_c$ and $\Delta\psi_{22}$ would arise due to inclinations or unequal masses (as shown in the previous case of quasi-circular binaries). In addition, the approximant \texttt{EccentricTD} is limited to systems with eccentricity up to $e=0.4$, so lensed events with higher eccentricity were not shown here but are expected to have larger deviations from unlensed GR waveforms. Therefore, we conclude that the approximate degeneracies would get only worse for more realistic eccentric waveforms.

\section{Template matching of lensed GWs}
\label{sec:SNR}

We have seen that, in general, lensing can modify the GW signal, with distortions possibly going beyond the general relativity (GR) predictions without lensing. It is then crucial to understand if these differences could be large enough to make a lensed signal to be missed in a GW search campaign and perhaps later, identified as a deviation from GR. From the analysis of the previous section, we know that Type III images simply have an overall sign difference with respect to the unlensed signal and will never be missed. Therefore, in this section we focus on analyzing Type II images.

We can assess these questions quantifying the matching of a lensed signal in a template search. Different statistics may be used to achieve this, and here we analyze the matched filter signal-to-noise ratio (SNR) of a single detector. We find that for a wide range of parameters of the binary, the SNR loss due to using unlensed GR templates in a GR lensed signal is at most a few $\%$.\footnote{Roughly speaking, a systematic SNR loss of $x\%$ implies that $1-(1-x)^3$ of the events will be lost since the detector horizon will shrink. Namely, a $5\%$ mismatch will lead to $\sim15\%$ of the signals missed.} 
Extreme cases of precession or eccentricity cannot be analyzed with the waveform approximants used here, and thus it is not possible to quantify if they will be detected.

The SNR of a signal $s(t)=h(t)+n(t)$, composed of a GW $h$ and noise $n$, with respect to a template $h_T(t)$ is given by \cite[Section 7.3]{Maggiore:1900zz}\footnote{Note that we are setting $t=0$ as the time of arrival of the GW.}
\begin{equation}
    \rho= \frac{^{~~}( s\vert h_T )^{~~}}{\sqrt{(h_T\vert h_T)}}\approx\frac{^{~~}( h\vert h_T )^{~~}}{\sqrt{(h_T\vert h_T)}}\,,
\end{equation}
where we are neglecting the correlation of the noise and the template and we have defined the inner product in Fourier space as:
\begin{equation}
    (a\vert b)=4\text{Re}\left[\int_0^\infty df \frac{\tilde{a}(f)\cdot \tilde{b}^*(f)}{S_n(f)}\right]\,,
\end{equation}
where tilded functions are in Fourier space and $S_n(f)$ is the single-sided power spectral density. The optimal SNR is achieved when the template matches the strain $h_T\propto h$, thus obtaining
\begin{equation}
    \rho_\text{opt}= \sqrt{( h\vert h )}\,.
\end{equation}
One should note that the matched filter SNR is insensitive to rescalings of the amplitude of the template.  

\begin{figure*}[t!]
\centering
\includegraphics[width = 0.95\textwidth]{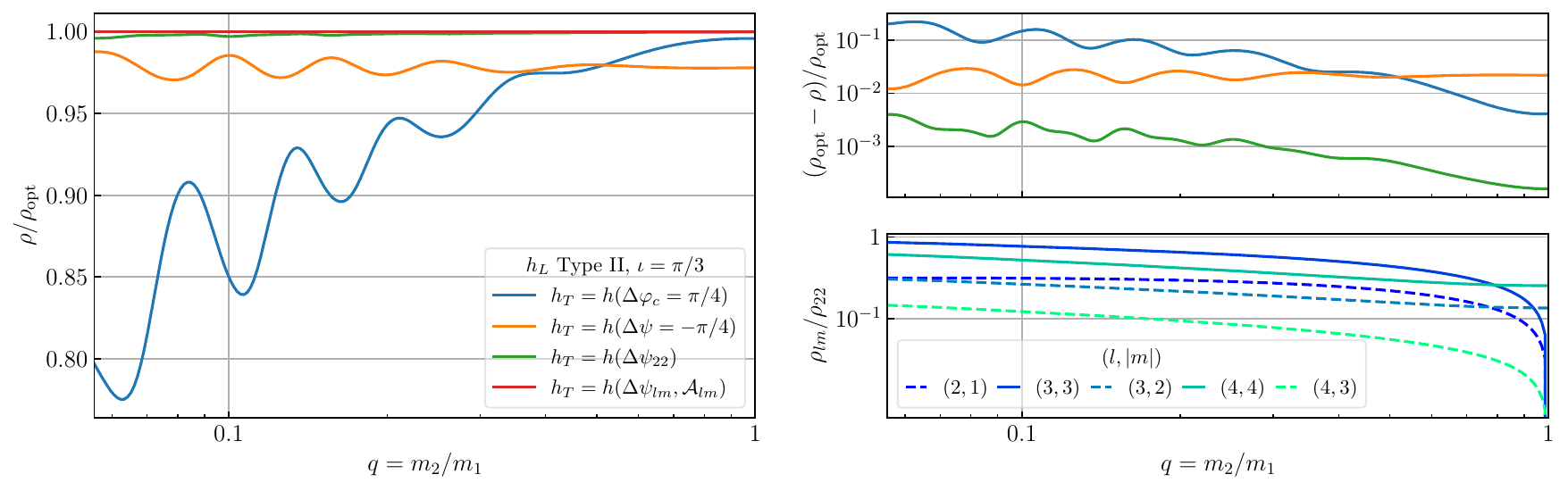}
 \caption{Matched filter identification of a lensed GW from a planar circular binary orbit. 
 On the left, we present the ratio of the matched filter SNR $\rho$ and the optimal SNR $\rho_\text{opt}$ for a type II GW image using different templates. On the top right panel we zoom into their relative difference. We use as templates an unlensed GW with the same parameters of the binary expect for a fixed shift in the coalescence phase $\Delta\varphi_c$, a fixed shift in the orientation $\Delta\psi$, a global source position/inclination dependent $(\theta,\phi,\psi,\iota)$ orientation  $\Delta\psi_{22}$ and amplitude change $\mathcal{A}_{22}$
 and a mode dependent orientation/amplitude change $\Delta\psi_{lm}\,,\  \mathcal{A}_{lm}$. Both $\Delta \psi_{22}$ and $\Delta \psi_{lm}$ are detector dependent quantities. 
 We present these quantities as a function of the mass ratio of the binary $q$. 
 In order to highlight the presence of higher modes, we quantify their individual SNR contribution in the lower right plot. 
 In all the panels we have chosen $\mathcal{M}_c=26.1 M_\odot$, $z_S=0.5$, $M_L=10^4M_\odot$, $z_L=0.1$, $\theta=0.3$, $\phi=0.4$, $\psi=1.5$ and $\iota=\pi/3$. The unlensed waveforms have been computed using \texttt{IMRPhenomHM} \cite{PhysRevLett.120.161102} and the SNR is computed against the O3 sensitivity curve.}
 \label{fig:snr_diff}
\end{figure*}

In the following we investigate the effect of higher modes, precession and eccentricity in the matched filter SNR. Details on the waveform approximants used can be found in Appendix \ref{app:waveform}. In our SNR computations we use the public estimate of the sensitivity curves during O3 described in \cite{Aasi:2013wya}, which can be found at \cite{sensitivity_curves_ligo}. 
It is to be noted that actual GW search pipelines determine the detection of a signal through a re-weighted SNR $\hat{\rho}$ which takes into account the matching of the templates to the signal in different frequency bins via a reduced $\chi^2$ \cite{Allen:2004gu,Babak:2012zx}. 
If the signal's time-frequency evolution does not match with the template, the $\chi^2$ re-weighting of the SNR employed by LIGO-Virgo \cite{Colaboration:2011np}
could further down-weight the overall significance of a detection. 
This is a generic feature of current search pipelines that will similarly affect lensed signals.  
The simple SNR results above are instructive, but only an implementation of a full search pipeline can address the resulting degradation in the strength of the detected signal, and whether lensed sources might be missed. We leave a thorough pipeline study for future work. 
Finally, we emphasize that the mismatch between a lensed signal and unlensed GR templates can also be used to identify the signal as strongly lensed, especially in the case of high SNR events.

\subsection{Nearly circular binaries}

In Sec.~\ref{sec:circular}, we found that for quadrupolar radiation  (or equivalently the 22 mode), a lensed GW is degenerate with a shift of the unlensed signal in the coalescence phase $\Delta\varphi_c$ or a (geometry dependent) change in the orientation angle given by $\Delta\psi_{22}$. Higher modes with $m\ne 2$ however break this degeneracy.

If only $m=\ell$ modes were present, a global $\Delta\psi_{22}$ change (together with a corresponding rescaling in amplitude) would give a nearly exact degeneracy. In general though, we expect to have contributions of $m\not=\ell$ modes, in which case some of the lensed images will be distorted with respect to the unlensed one, and will not conform exactly to GR waveforms (i.e.~for a given GR unlensed waveform, there is no shift in astrophysical parameters that reproduces the lensed waveforms). If those modes are subdominant, the $\Delta\psi_{22}$ shift would still be a good approximate degeneracy. 

It is important to note that one could always construct an exact template for the type II image by phase shifting the type I image in the frequency domain. This would be the most practical approach when dealing with signals dominated by the higher modes. As a pedagogical exercise, we can also think of other ways to construct a degenerate template. In particular, one way of building an almost exact template for non-precessing binaries with higher modes is to change the orientation angle and amplitude independently for each mode. In this case, the degeneracy would be given by a shift $\Delta\psi_{\ell m}$ and a change of the amplitude $\mathcal A_{\ell m}$ following equations (\ref{ChivsAngles})-(\ref{AvsAngles}). Alternatively, one could also define a ($\ell$,$m$) dependent coalescence phase. 
It is to be noted that, in any case, this new template will no longer describe a standard GR waveform.

We have quantified the degree of precision of these different degeneracies by computing the matching between the template and lensed signal. This analysis determines whether standard searches using GR waveforms (namely $\Delta\varphi_c$ and $\Delta\psi_{22}$ shifted templates) will detect or miss real lensed signals. We summarize these results in Fig.~\ref{fig:snr_diff}, where we consider a non-spinning binary with inclination $\iota=\pi/3$ (and rest of lens and binary parameters as in the previous section), and analyze the matched filter SNR of the type-II lensed image as a function of the mass ratio. On the left panel we present the ratio of the SNR with respect to the optimal SNR. The templates used correspond to the type-I waveform (that is equivalent to the unlensed waveform) with additional shifts in the astrophysical parameters, whereas the optimal template uses the type-II lensed waveform itself. 
For equal masses we can see how the fixed orientation shift $\Delta\psi=-\pi/4$ gives the worst fit. The coalescence shift $\Delta\varphi_c=\pi/4$ gives a better match although not exact due to the presence of higher modes. The $\Delta\psi_{22}$ shift gives such a good match that it cannot be distinguished in this panel. Lastly, the mode dependent shift $\Delta\psi_{\ell m}$ gives an exact (to numerical accuracy) match for any mass ratio $q$. In order to distinguish better the difference at small $q$, we plot in the upper right panel the relative difference of the SNR. There, it is clear that $\Delta\psi_{22}$ gives the best fit (apart from $\Delta\psi_{\ell m}$, which was constructed to exactly match the lensed image and thus does not itself appear in this plot since the relative difference is 0). It is clear from this panel also that the error in the other matches increases as $q$ decreases. This can be explained from the lower right panel where we show the SNR of each individual higher mode included in the waveform model. One should note that $m=\ell$ modes always dominate which explains why $\Delta\psi_{22}$ gives a very good approximation to the type-II lensed image.

Altogether, we find that the are GR templates that mimic to very good approximation the lensed type II image of asymmetric mass binaries containing higher modes. This happens due to the approximate degeneracy of lensed signals with a (detector and inclination dependent) change in the orientation angle.  
For the range of mass ratios that we have explored (which is limited by the calibration of the waveform approximants) relative SNR differences using $\Delta\psi_{22}$ will be below $1\%$. This means that one will not be missing non-precessing, circular lensed events with standard searches. 
We have tested that these results hold for the range of masses relevant to advanced LIGO and Virgo, $\mathcal{M}_{z}\sim[2,200]M_\odot$. One should note that the relative amplitude of higher modes is sensitive to the mass ratio $q$, but not to the total mass. However, for higher mass binaries the merger frequency of the dominant quadrupole radiation might lie beyond the optimal sensitivity of the detector, making the higher modes $m\ne 2$ to contribute more to the total SNR \cite{Mills:2020thr}. This in turn makes $\Delta\varphi_c$ to perform poorly, but $\Delta\psi_{22}$ remains a good approximation as we have checked. 
Lensed binaries with extreme mass ratios could still be detected by constructing targeted templates that phase shift the type I image in the frequency domain.

\subsection{Precession}

As shown in \ref{sec:precession}, type II images of precessing binaries are more difficult to model with standard waveforms since the time dependent nature of the orbit's precession prevents it from having a constant orientation angle shift. This is also true for the (non-GR) $\Delta \psi_{\ell m}$ template that we constructed in the previous section by a mode-dependent change in the orientation angle and amplitude.

\begin{figure}[t!]
\centering
\includegraphics[width = 0.95\columnwidth]{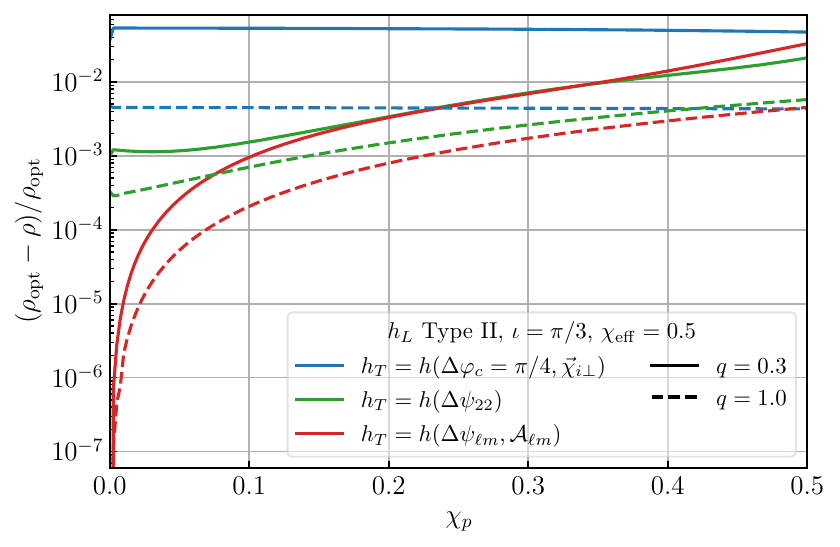}
 \caption{Matched filter identification of a lensed GW with precession. 
 The  SNR differences between the matched filter SNR $\rho$ and the optimal SNR $\rho_\text{opt}$ for a type II GW image
 use as templates the curves constructed as in Fig.~\ref{fig:wave-form_precession}
 but with varying spin precessing parameter $\chi_p$ defined in (\ref{eq:chi_p}). 
 We compute waveforms using \texttt{IMRPhenomPv3HM} 
 \cite{Khan:2019kot}, which contains higher modes in the co-precessing frame, and use the same binary parameters of Fig.~\ref{fig:snr_diff} fixing $\chi_\text{eff}=0.5$.}
 \label{fig:snr_precession}
\end{figure}

We quantify the template matching of GR ($\Delta\varphi_c$ and $\Delta\psi_{22}$) and non-GR waveforms ($\Delta\psi_{\ell m}$) with lensed precessing binaries in Fig.~\ref{fig:snr_precession}, where we consider two cases of binaries with equal and unequal masses, both having inclination $\iota=\pi/3$ and $\chi_\text{eff}=0.5$.
We show the relative SNR as a function of the precession parameter $\chi_p$, see Eq.~(\ref{eq:chi_p}). Recall that we are using a waveform model including higher modes in the co-precessing frame. Thus, in the absence of precession only the $\Delta\psi_{\ell m}$ template gives an almost perfect match. 
The coalescence phase shift supplemented by a rotation of the spins in the orbital plane gives an exact (to numerical accuracy) fit when there is only quadrupole radiation in the co-precessing frame. In this more general case, we can see that there is always a similar SNR difference. In terms of comparing $\Delta\psi_{22}$ and $\Delta\psi_{\ell m}$ we can see that the expected hierarchy is preserved for small $\chi_p$. However, for larger precession parameters ($\chi_p\gtrsim 0.2$ in this example) both behave similarly. As previously discussed, a template search would find a better $\Delta\psi$ that corrects for the average change in the viewing angle, but still would not perform as well as the non precessing case due to its time dependence.
In all the cases we observe that a smaller mass ratio (solid lines correspond to $q=0.3$ and dashed ones to $q=1$) gives a larger mismatch. 

One should note that the results presented in Fig.~\ref{fig:snr_precession} are subject to the choice of the individual spin vectors $\vec{S}_{1,2}$. For this example we have chosen the simplest mapping between $\vec{S}_{1,2}$ and $\chi_\text{eff},\chi_p$, described in the previous section. We have checked that other choices have a similar SNR trend. As in the previous analysis of higher modes, our parameter space search is limited by the calibration of the waveform models we are using. In this case, the dimensionless spin magnitudes have to be below $0.5$.

We conclude that in this range of precession parameters $\chi_p$ lensed images of precessing, circular binaries would not be missed with a standard template, as the error, although larger than before, is still less that $5\%$ for $\Delta\psi_{22}$.  
The tendency of Fig.~\ref{fig:snr_precession} seems to suggest that more extreme scenarios precessing faster may be missed with standard GR templates, although this goes beyond the limit of our waveform approximants. We have tested though that the above results hold for the relevant mass range of present GW interferometers. 
Again, for extreme precessing binaries an exact lensed template could be built by phase shifting the earliest image in frequency space.

\subsection{Eccentricity}

\begin{figure}[t!]
\centering
\includegraphics[width = 0.95\columnwidth]{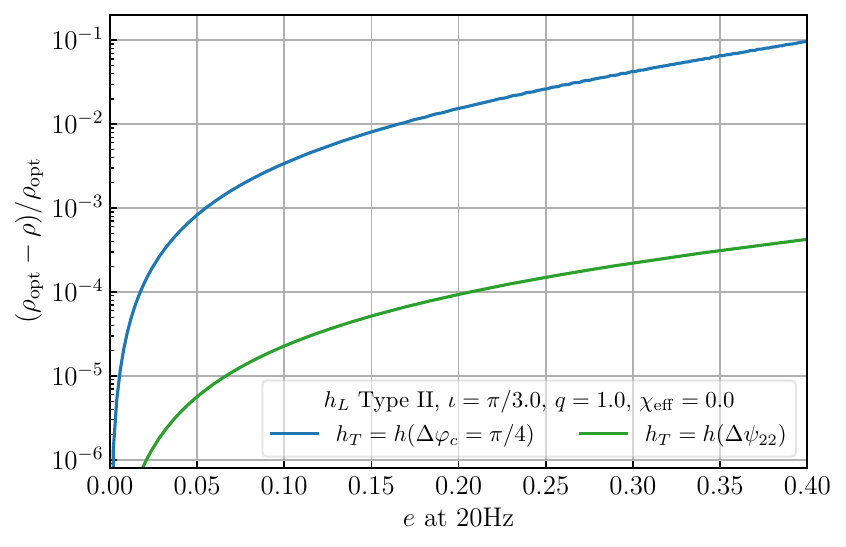}
 \caption{Matched filter identification of a lensed GW with an eccentric orbit. 
 The SNR difference between the matched filter SNR $\rho$ and the optimal SNR $\rho_\text{opt}$ for a type II GW image using as templates the curves described in Fig.~\ref{fig:wave-form_eccentricity} but as a function of eccentricity. 
 We compute the SNR of the waveform inspiral using \texttt{EccentricFD}, valid for the quadrupolar radiation of a non-spinning binary up to the inner most stable circular orbit \cite{Huerta:2014eca}. We use the same binary parameters of Fig.~\ref{fig:snr_diff}, fixing $q=1$, $\iota=\pi/3$ and $\chi_\text{eff}=0$. The eccentricity is fixed at the initial frequency of 20 Hz. The SNR is computed against O3 sensitivity.}
\label{fig:snr_eccentricity}
\end{figure}

As we have discussed in Sec.~\ref{sec:eccentricity}, even for quadrupole-only radiation we will have deviations from the match of the type II lensed signal that is provided by $\Delta\varphi_c=\pi/4$ or $\Delta\psi_{22}$.
We quantify this result in Fig.~\ref{fig:snr_eccentricity}, where we plot the relative SNR difference as a function of the eccentricity for a non-spinning, equal-mass binary. We use a waveform model that only includes the $\ell=2$ modes and for that reason we only present the $\Delta\varphi_c=\pi/4$ and $\Delta\psi_{22}$ curves. 
The SNR is computed up to the inner most stable circular orbit since that is the limiting frequency of the \texttt{EccentricFD} waveform model \cite{Huerta:2014eca}. 
Moreover, we restrict to equal mass binaries to reduce the error of not including higher modes. We check that when $e\to 0$ the degeneracy is recovered. As $e$ grows, the relative difference becomes larger. 
We fix the initial eccentricity at 20Hz. The mismatch is larger when it is normalized at a higher frequency since the eccentricity evolves with time and hence frequency.
We have checked that in the regime of eccentricities considered here the eccentricity evolution can be well approximated by $e(f)\sim e_0 (f/f_0)^{-19/18}$ \cite{Yunes:2009yz}, allowing to extrapolate our results to other reference frequencies.

Similar to the analysis of Fig.~\ref{fig:snr_diff}, we find that the $\Delta\psi_{22}$ shift gives a better fit. For our simple example and the range of eccentricities explored, the differences are small for $\Delta\psi_{22}$, less than $1\%$ at both O3 (as plotted in Fig.~\ref{fig:snr_eccentricity}) and advanced LIGO design sensitivity. 
For $\Delta\varphi_c=\pi/4$ the mismatch can grow up to $10\%$ for $e=0.4$ at 20Hz with O3 sensitivity.  
Again, our parameter exploration is limited by the calibration of the waveform to $e\leq0.4$, but has covered a wide range of masses. Altogether, although our analysis only includes the inspiral phase and does not account for higher modes, given the small differences observed we conclude that most likely no eccentric lensed events would be missed with second generation ground based detectors. Future third generation detectors and space-based observatories like {\it LISA} will be more sensitive to the eccentricity and might find larger differences.

\section{Multiple image searches}\label{sec:waveform}

As we have discussed in previous sections, strong lensing in the geometric optics regime produces multiple images. In general a lensed GW image may not be consistent with standard general relativity, as lensing in the stationary-phase approximation will cause frequency-dependent phase shifts, generating what naively will appear to be non-GR waveforms. However, as shown in the previous section, to a good approximation the images differ mainly in their amplitude, arrival time, and phase. 
Therefore, except for the luminosity distance, coalescence time, and binary phase parameters (coalescence phase or orientation angle), the rest of the parameters describing the binary (such as detector frame masses, localization of the source on the sky, inclination, mass ratio, etc.) remain the same (recall that deflection angles changing the sky position or polarizations are beyond present/future GW detector's capabilities as summarized in Appendix~\ref{sec:pol}). As a consequence, analyses in the literature aiming at determining if a set of detected events could result from strong lensing of the same source typically consist of two steps: lensed pairs identification and a subsequent joint parameter estimation (PE) consistency test. Searches of lensed GWs have been performed in \cite{Haris:2018vmn,Li:2019osa,McIsaac:2019use,Hannuksela:2019kle, Broadhurst:2019ijv, Dai:2020tpj}. In this section, we present how these previously proposed search strategies can also include information on the phase to evaluate the lensing hypothesis.

Based on our results of the previous section, an efficient search for strongly-lensed GW sources would be achieved by including:
\begin{enumerate}
\item \emph{Phase informed lensing candidates:}  The effects of lensing will leave the sky position, masses, and spins of the sources of unchanged. The polarization state is also effectively unchanged for realistic astrophysical lenses. Magnification will alter the inferred distances, and lensing will also impact the phase of the waveforms as discussed in previous sections. 
The inclusion of phase information in the analysis of potential lensed candidates would reduce the false-alarm rate and increase confidence in lensing identifications. In particular, the difference in phase between the type I and II, and type I and III images should be $\pi/2$ and $\pi$, respectively.

Once a catalog of sources has been assembled, a search can be performed identifying pairs (or higher numbers) of sources which have consistent localization, mass, and phase. Additional quantities, such as spins, can also be checked for consistency (see e.g.~\cite{Hannuksela:2019kle}), although in practice these other parameters are more poorly measured and have less constraining power. Since the $\ell=m=2$ modes generally dominate the waveforms, the phase relations are often well approximated by a shift of the coalescence phase $\varphi_c$ or orientation angle $\psi$ and amplitude $\cal{A}$.  
Catalogs of GW sources, such as GWTC-1 \cite{LIGOScientific:2018mvr}, are often accompanied by full parameter estimation results for catalog entries. It is therefore possible to consider exploring 
approximate, partial degeneracies between $\psi$ and $\varphi_c$. In particular, for type II images, a conservative selection criteria is that either the point ($
\Delta\psi =\Delta\psi_{22}$, $\Delta \varphi_c=0$)\footnote{It is to be noted that $\Delta\psi_{22}$ is a detector and inclination dependent quantity, but standard parameter estimation data releases are given in a common Earth-fixed frame. We have verified that, except for edge-on binaries, the actual value of $\Delta\psi_{22}$ is not very sensitive to the detector. However, an improved application would incorporate the effects due to different detectors in evaluating $\Delta\psi_{22}$.}
or ($\Delta\psi =0$, $\Delta \varphi_c=\pi/4$) lies within the allowed region of the joint posterior of the $\psi$ and $\varphi_c$ parameters, when marginalizing over possible additional free parameters.\footnote{This test assumes that the wave-form distortions induced by the phase shift of the type II image do not affect significantly estimates of other intrinsic parameters
 compared to the type I image. Possible systematic effects in the parameter estimation of type II images with $\Delta\psi_{22}$ 
should be explored in the future.} For type III images, the lensed signal will always be degenerate with an unlensed signal with an orientation shift of $\pi/2$.

It is to be noted that type II images of events with extreme precession or eccentricity would need to be treated separately, since for these sources the phase relations between primary and lensed images are no longer dominated by a simple phase shift of the $\ell=m=2$ mode. It is apparent from current catalogs that high precession/eccentricity sources do not constitute a large fraction of detected sources. Given their scarcity, it is likely that cases of strongly-lensed high precession/eccentricity sources could be straightforwardly identified in the catalogs and identified as interesting candidates for lensing. 
Once pairs (or higher numbers) of sources have been identified as potentially interesting lensing candidates, a subsequent joint parameter-estimation step is helpful to further constrain the parameters of the source. 

We note, however, that in the future, when a network of ground-based detectors is able to detect independently each polarization and precisely measure the orientation angle $\psi$, the approximate degeneracy in $\Delta\psi_{22}$ should not be used as a selection criteria.

\item \emph{Phase shifted waveform pair analysis:} once candidate lensing event have been identified in the catalog, joint parameter estimation can be performed under the assumption that the events are indeed images of the same source. An example of a joint analysis of lensed images was performed in \cite{Seto:2003iw}, although without taking into consideration phase shifts due to lensing. 
Here we propose performing parameter estimation which directly accounts for possible phase shifts due to lensing, for any compact binary source. 
In the frequency domain the phase shift corresponds to multiplying the
template waveform by a constant:
\begin{align}
\label{eq:shift1}
    &\text{Type II:} & &\tilde{h}_\text{II}(\omega)= e^{-i\frac{\pi}{2}} \tilde{h}_\text{I}(\omega) \,,\\
    \label{eq:shift2}
    &\text{Type III:} & &\tilde{h}_\text{III}(\omega)= e^{-i\pi} \tilde{h}_\text{I}(\omega) \,,
\end{align}
for $\omega>0$, with $\omega<0$ defined by conjugation that preserves the real part of the waveform. Parameter estimation could then be performed using both unlensed and lensed waveform templates.

This approach allows for a range of additional consistency tests.
For instance, the most basic test would be to promote the relative phase $\pi/2$ and $\pi$ to a free parameter $\alpha$, and check if the posterior distribution is consistent with the lensing values $\alpha=0,\pi/2,\pi$. This is a generalization of the approach pursued in~\cite{Dai:2020tpj}, where consistency was tested with the coalescence phase of the time domain signals. 
One can generalize this phase consistency test to any combination of astrophysical
parameters, such as the mass ratio, $\psi$, or $\varphi_c$, to test the lensing 
hypothesis more fully.

\item \emph{Targeted multiple image and subthreshold search:} for those candidate lensing pairs which pass the previous steps, the last step would be to test whether any additional candidates, above or below threshold, are consistent with being additional images of the same source.   This would follow the same procedure of the previous step and test joint consistency of the phase and astrophysical parameters with those of the original pair. This approach was pursued in~\cite{Dai:2020tpj} (again using $\varphi_c$ as a reference), and unearthed a sub-threshold trigger, GWC170620, as a potentially associated image.
We note that, since the parameters of the source are determined in steps~1 and~2 above, the search for additional candidates can be improved. Instead of a generic search, the space of interesting events is severely constrained to sources which are consistent with the parameters determined above, modulo magnification (i.e.~different luminosity distances) and phase shifts. When applied to raw data, this targeted search allows for the confident identification of sources which otherwise might be lost in noise. 

\end{enumerate}

Here we have only described how to incorporate phase information on a lensing consistency analysis between multiple images. Nevertheless, consistency of waveforms under the lensing hypothesis is only one aspect of estimating the probability of lensing. In future work it is also important to address the astrophysical priors, such as the expected probability of strong lensing, and the expected distribution of lensing configurations with their time-ordering of the arrival of various image types. For example, the expected incidence of strong lensing in the existing LIGO/Virgo sample is low \cite{Li:2018prc,Oguri:2018muv}; the strong lensing rate at $z\sim 1$ is expected to be less than $1\%$, and the detected population is consistent with being at low redshift~(e.g.~see~\cite{2018ApJ...863L..41F,2019ApJ...882L..24A}; although see \cite{Broadhurst:2020moy} for an alternate view).
In addition to the incidence of lensing, the morphology of the lens can also provide insight. As it is noted in \cite{Dai:2020tpj}, the lensing configuration they propose with a total of 7 images, and a bright type III image, would be a highly unusual system. This therefore argues against the lensing hypothesis. Similarly, existing electromagnetic lensing surveys, such as those identifying multiply-imaged quasars, should be an excellent source of information for determining the expected lensing rate, image types, and time ordering.
Finally, it is important to consider the false-alarm rate from the GW data perspective. Given the elongated shapes of GW sky localizations \cite{2016ApJ...829L..15S}, the uneven mass distribution of detections with a significant fraction of sources at high mass~
\cite{2017ApJ...851L..25F,2020ApJ...891L..31F}, and significant errors on inferred binary parameters such as total mass, spin, and phase, a large GW catalog is likely to find sources consistent with lensing even if no strong lensing is present. It is important to account for this false-alarm probability in any evaluation of the likelihood of lensing. 
This can be achieved by performing an injection campaign.

It is also to be noted that if multiple GW events are confirmed to be strongly-lensed images of the same source, it may be possible to identify the host galaxy \cite{Hannuksela:2020xor}. This is because the host galaxy is likely to also be multiply imaged, and this could be searched for in the GW localization region. 
Electromagnetic observation of the lensed host galaxy might also help confirm the lensing hypothesis, and further constrain the lensing geometry. If electromagnetic localization is achieved, 
the GW source sky localization greatly improves \cite{Hannuksela:2020xor}, and these strongly-lensed events can be used 
to test cosmology, in an analogous way to lensed quasars \cite{Wong:2019kwg}.

\section{Discussion}\label{sec:discussion}

Strong gravitational lensing of GWs has the potential to provide new avenues to test the underlying theory of gravity, the cosmological expansion history, and the distribution of matter in the universe (see e.g.~\cite{Oguri:2019fix}). The identification of strong lensing of gravitational-wave sources necessitates an understanding of the characteristics of the lensed images, and an implementation of these lensed waveforms in the lensing search pipelines.

In this work we have explored the relative phase of lensed GWs as a key indicator of strong lensing. We have demonstrated that  lensed waveforms can be indistinguishable from unlensed ones with shifted astrophysical parameters. The respective $\pi/2$ and $\pi$ phase shift of type II and III images can be mimicked in different ways. In particular, we have found that a detector and inclination dependent shift of the orientation angle $\Delta\psi_{22}$ (see Eq.~(\ref{ChivsAngles})), provides a very good approximate template for lensed images. The template is not always exact due to the following effects:
\begin{enumerate}[label=\emph{\roman*)}]
\item the presence of higher order modes with  $ m \ne \ell$,
\item the precession of the orbital plane,
\item the ellipticity of the orbit,
\end{enumerate}
which cause the lensed GW signals to deviate from GR waveforms. We also found that the lensed images can be degenerate with a shift in the coalescence phase $\Delta \varphi_c=\pi/4$. This latter degeneracy is less general since it is already broken when $m\ne 2$ modes are relevant. We note, however, that for the set of eccentric binaries that we have explored, these approximate degeneracies are preserved to a very high degree.

We emphasize that lensed waveforms may not correspond to standard general relativity waveform templates. 
As shown above, in the case of gravitational lensing of gravitational-wave sources, individual modes may suffer phase shifts, and thus the final waveform may be altered. A naive analysis might thus claim deviations from general relativity, when instead the source has been strongly lensed. These deviations will need to be incorporated in future Testing GR analyses~\cite{2019PhRvD.100j4036A}, so that lensed sources are not erroneously identified as indicating a breakdown of general relativity. This differs from intuition based on the case of strong lensing of electromagnetic sources, where in general only the intensity is measured (averaged over times long compared with the frequency), and thus phase effects become irrelevant. 
Interestingly the possible distortion of strongly lensed GWs also opens the possibility to identify strong lensing without the need to detect multiple images.

We have studied a large sector of the compact binary parameter space with mass ratios above $q=1/18$, precession parameter below $\chi_p=0.5$ and eccentricities below $e=0.4$ at 20Hz. We find that, for a single detector, lensed GWs with parameters in this range will likely not be missed by standard GR templates, as the signal-to-noise ratio loss due to waveform mismatch will be smaller than $1\%$ for asymmetric binaries and $5\%$ for precessing and eccentric binaries when using a detector-dependent orientation angle shift. When shifting instead the coalescence phase the mismatch is lower than $10\%$ for parameters in these ranges. 
More extreme choices of the precession or eccentricity may potentially produce larger mismatches that could lead to missing the signal with a traditional template bank search. In addition, when having multiple detectors, a detector-dependent phase shift may further decrease the detection confidence of an event, as current searches only allow for detector-independent parameters and even test for their consistency. Generating a targeted template bank (see Eq.~(\ref{eq:shift1}) and (\ref{eq:shift2})) will overcome these problems, although the probability of detecting these rare binaries seems low at the moment with present GW detectors. 

Even with a single image, whenever the lensed signal differs form GR templates, in principle the waveform alone would demonstrate that the source is multiply imaged by gravitational lensing.
As shown after the submission of this work, high signal-to-noise detections by 3G instruments could indeed allow the direct identification of type II images through their waveform distortions \cite{Wang:2021kzt}.
Furthermore, this offers an important consistency test of strong lensing, since the waveforms provide a direct signature of lensing. This is a unique aspect of lensing of GW sources, one that is absent in electromagnetic sources because they are not phase coherent. 

Even though most lensed events will not be missed if phase shifts of templates are not taken into account, 
we advocate for the inclusion of phase shifts as a criteria to assess the probability that a given set of events are lensed images from the same source. We outline our optimal search strategy that incorporates information on the phase. This starts with selecting candidates which overlap in their sky positions and masses, and are consistent with the approximate degeneracy in the coalescence phase or orientation angle. 
Moreover, we propose that the joint parameter estimation analysis of multiple events should be performed with the inclusion of the phase shifted signals in the frequency domain. In this way, the analysis will be exact for any type of binary, and consistency of the phase and source parameters could serve as a consistency test for the lensing hypothesis in one or more dimensions.

Upon completion of this work, ref.~\cite{Dai:2020tpj} analyzed a set of lensed candidate GW events in the second observing run of LIGO-Virgo, by taking the phase shifts studied here as changes in the coalescence phase. 
Without entering into a discussion of the probabilities of this being an actual lensing system, we find it encouraging  
that the information from the phase shift could help in narrowing down and identifying lensing candidates. The analysis of \cite{Dai:2020tpj} was restricted to waveforms with (22)-modes only, 
where  $\Delta \varphi_c$ indeed mimics the phase shift and may be used as a lensing consistency test.

Looking to the future, a larger network of ground-based GW detectors with improved sensitivities will boost the search for strongly-lensed GW events. The main improvements will be due to the more precise sky localization and the possibility of independently determining GW polarization. The latter would provide a new consistency test, since lensing should change the polarization state negligibly, and thus distinguish between orientation and phase shifts. For instance, a test of the number of GR polarizations with multiple lensed images has been proposed in \cite{Goyal:2020bkm}. 
In addition, this phase consistency test could be a key lensing discriminator for the space-based interferometer {\it LISA}, since the signals of high-redshift super-massive black-hole binaries could be in band from days to months, providing an exquisite measurement of the phase. 
Note that even for a single detector, a source of long-lived in band wave emission will change its relative location in the sky and polarization as the detector moves over the course of observation. These effects would allow one to determine the source location and independent polarizations of the wave, in which case a shift in the orientation angle would no longer be degenerate with a lensed phase shift. 
Finally, we emphasize that third-generation GW detectors are expected to detect events up to redshift $z\sim 10$, in which case the lensing probability will increase considerably and therefore a comprehensive lensing analysis will be necessary. 

\acknowledgements
We are grateful to the LVC lensing group for their feedback, to Otto Hannuksela and (Rico) Ka-Lok Lo for their comments on the draft, as well as to Reed Essick and Zoheyr Doctor for their valuable comments, and Fei Xu for her valuable feedback on the code developed for this project.   
J.M.E.~was supported by NASA through the NASA Hubble Fellowship grant HST-HF2-51435.001-A awarded by the Space Telescope Science Institute, which is operated by the Association of Universities for Research in Astronomy, Inc., for NASA, under contract NAS5-26555. 
D.E.H.~was supported by NSF grants PHY-1708081 and PHY-2011997. D.E.H. also gratefully acknowledges support from the Marion and Stuart Rice Award.
W.H.~was supported by the U.S.~Dept.~of Energy contract DE-FG02-13ER41958 and the Simons Foundation. 
R.M.W.~was supported by NSF grant PHY18-04216 to the University of Chicago.
The authors were supported by the Kavli Institute for Cosmological Physics at the University of Chicago through an endowment from the Kavli Foundation and its founder Fred Kavli.

\appendix

\section{Polarization}\label{sec:pol}
In the previous discussions we have 
assumed that lensing does not change the polarization state of the GW signal which is, in principle, a useful additional consistency check on lensed signals. Here we examine the extent to which the polarization can differ between the signals.

In geometric optics, the polarization is parallel-propagated along the path of a given image. Therefore, we expect the polarization to change from emission to detection by an amount given by the deflection angle \cite{Hou:2019wdg}. In particular, if the source-lens-observer are located in the $z-x$ plane, then lensing will lead to a rotation of $|\alpha|$ in the propagation direction (and thus the polarization plane) around the $y$ axis. As shown explicitly in \cite{Hou:2019wdg}, this change in direction can be reinterpreted as a change in the polarization of the detected wave with respect to the initial propagation direction, which leads to different effective tensor and vector modes. 

If we now consider two images, in addition to the individual changes in direction of polarizations, we will also have a different emission angle for both images. In this case, both images are effectively seeing the source from different perspectives and hence a change in polarization between them will be present. This change in polarization will be due to a change in the inclination and polarization angles, as well as the mode structure of the signal as described in the main text. The observed angle is given by $\Delta x_{o}=x_{+}+x_{-}$, and then the emitted angle is given by:
\begin{equation}
    \Delta x_{e}= \Delta x_{o} - \left(\frac{\alpha_{+}}{\theta_E}+\frac{\alpha_{-}}{\theta_E}\right) =\Delta x_{o}\left(1-\frac{D_S}{D_{LS}}\right),
\end{equation}
where $\alpha$ are the deflection angles of both images. We thus generically have that $\Delta x_{e}\sim \Delta x_{o}= \sqrt{y^2+4}$ for a point lens. 
For a reference galaxy lens with mass of $M\sim 10^{12}M_{\odot}$ at Gpc distance, then the Einstein angle is of order:
\begin{equation}
    \theta_E\sim 1^{''}\sqrt{\frac{M}{10^{12}M_{\odot}}}\sqrt{\frac{1\text{Gpc/rad}}{D}},
\end{equation}
with $D$ the effective lensing distance. The larger the $y$ (source angular position) then the larger the emission angle. However, for typical strong lenses, the angular separations are of order arcseconds (e.g.~in lensed quasars \cite{1997ApJ...486...75M, 2019MNRAS.483.4242L}), corresponding to $y\lesssim \mathcal{O}(1)$. Therefore, we expect to have arcsec differences in the estimated angular parameters such as orientation, inclination, and position in sky.  Since the angular dependence on the polarization scales depends on the multipole moment or angular frequency as $2\pi /\ell$, for the low order modes that dominate GW signals, these induced changes are orders of magnitude below the current sensitivities, and hence we can safely neglect these effects.

\begin{figure*}[t!]
\centering
\includegraphics[width = 0.98\textwidth]{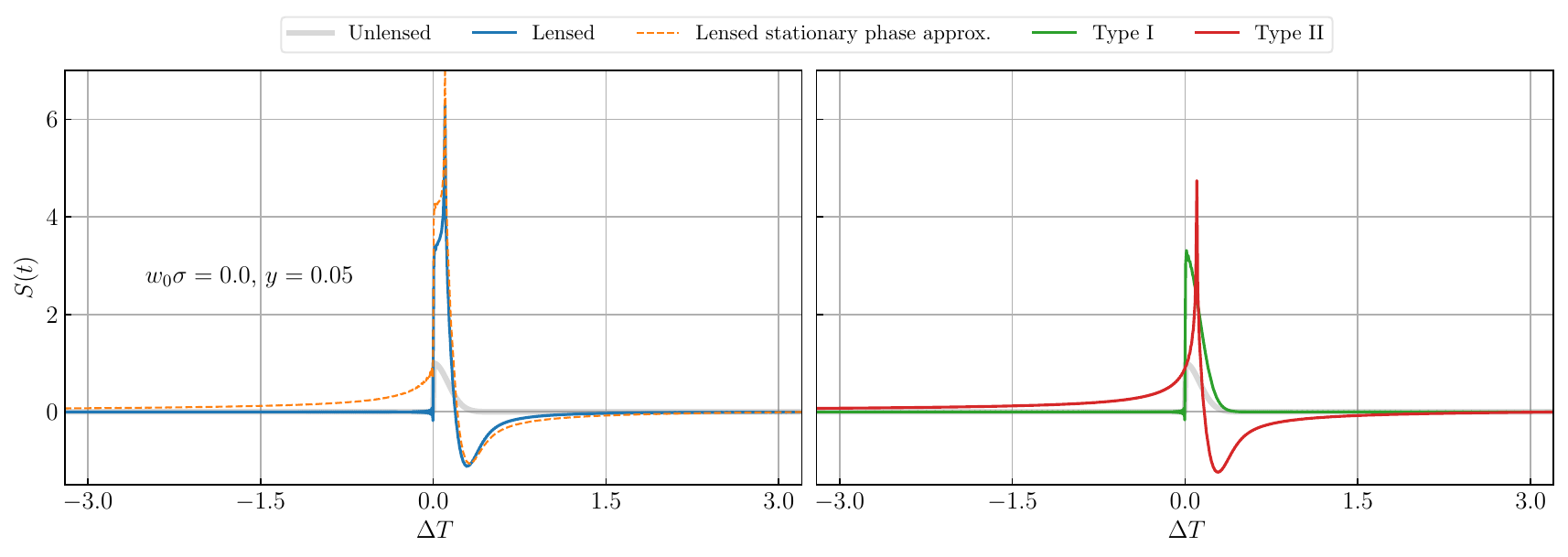}
 \caption{Effect of a point mass lens on a half Gaussian pulse where $S(t)\not=0$ for $t>0$, see Eq.~(\ref{eq:half_gaussian}). The stationary phase approximation breaks near the caustic at $y=0$, since it predicts an acausal tail at $\Delta T<0$. On the left we present the unlensed signal together with the lensed signal and the stationary phase approximation (dashed line) for $y=0.05$. On the right we plot the type I and type II images to show that the acausal tail originates from the latter.}
 \label{fig:acausal}
\end{figure*}

\section{Waveform approximants} \label{app:waveform}

In our analysis we generate GW signals using the phenomenological waveform family \texttt{Phenom} \cite{Ajith_2007} implemented in \texttt{LALsuite} \cite{lalsuite} calling the waveforms through \texttt{pyCBC} \cite{alex_nitz_2019_3546372}. We compute the higher modes waveform using \texttt{IMRPhenomHM} approximant \cite{PhysRevLett.120.161102}. This waveform model contains $(\ell m)=(22),(21),(33),(32),(44),(43)$ modes and uses the precursor \texttt
{IMRPhenomD} \cite{PhysRevD.93.044006} to compute the 22 modes, which has been calibrated with numerical relativity simulations in the ranges $1/18\leq q \leq 1$ and $\vert \chi_i\vert \leq 0.85$. 

To account for precession, we use the waveform approximant \texttt{IMRPhenomPv3HM} \cite{Khan:2019kot}. This is the latest of the family of waveforms \texttt{IMRPhenomP} \cite{PhysRevLett.113.151101}. This waveform class is constructed taking advantage of the fact that there exist a non-inertial frame (the co-precessing frame) approximately following the orbital plane where the effect of precession is minimized. Then, a precessing waveform can be modeled by tracking the time dependent rotation from the co-precessing frame to the radiation frame. \texttt{IMRPhenomPv3HM} includes the same higher modes that \texttt{IMRPhenomHM} in the co-precessing frame. This waveform model has been calibrated with numerical relativity simulations in the range $q\leq0.2$ and spin magnitudes $<0.5$. 

In order to test the behavior of the quadrupole radiation in a precessing scenario we revert to the older version \texttt{IMRPhenomPv3} \cite{Khan:2018fmp}. This approximant only computes the quadrupole moment (2,2) in the co-precessing frame. There are however other modes (with $\ell=2$) in the radiation frame due to mixing between $m$'s from rotation to the inertial frame. \texttt{IMRPhenomPv3} is calibrated with \texttt{IMRPhenomD}, extending the initial calibration of \texttt{IMRPhenomP} valid only in the range $1 \leq q \leq 3$ and $\vert\chi_\text{eff}\vert \lesssim 0.75$ as determined by \texttt{IMRPhenomC} \cite{PhysRevD.82.064016}.  
\texttt{IMRPhenomPv3} also extends \texttt{IMRPhenomPv2} \cite{PhysRevD.82.064016} by including a second spin effect.

Finally, to model the effect of eccentricity we use the waveforms \texttt{EccentricTD} and \texttt{EccentricFD} for the time and frequency domain respectively \cite{Huerta:2014eca}. This approximant is only valid for non-spinning binaries. It only includes the quadrupolar ($\ell=2$) radiation during the inspiral but accounts for higher order post-Newtonian corrections in the phase.  
It has been proven to be accurate in the range $e<0.4$.

\section{Breakdown of the stationary phase approximation near caustics}
\label{app:caustics}

In this appendix we take a closer look at the break down of the stationary phase approximation near caustics, as well as its impact on gravitational wave SNRs. 
This approximation breaks when the time delay between the images is shorter than the period of the wave, $\Delta t_d\cdot\omega\ll1$, and interference becomes a relevant phenomena. For a point mass, the time delay $\Delta t_d$ is a function of the source-lens alignment $y$ (see Eq.~(\ref{eq:Td})) and the mass of the lens (determined by the Schwarzschild  diameter crossing time $t_M$). We focus on the strong lensing regime, $y \lesssim 1$, 
and explore the limit of $y\to 0$ and $\omega\to 0$. 
As a pedagogical exercise, we begin with a simple toy example and then investigate how well standard (unlensed) waveform templates can match these lensed signals. 

As discussed in the main text, saddle point (type II) images predict a large tail of early arrivals that in the limit of source-lens alignment ($y\to0$) may lead to an apparent acausal behavior if that arrival precedes that of the type I image, where the time delay is a minimum.  This erroneous behavior indicates a waveform that differs from the stationary phase approximation. 

\begin{figure*}[t!]
\centering
\includegraphics[width = 0.95\textwidth]{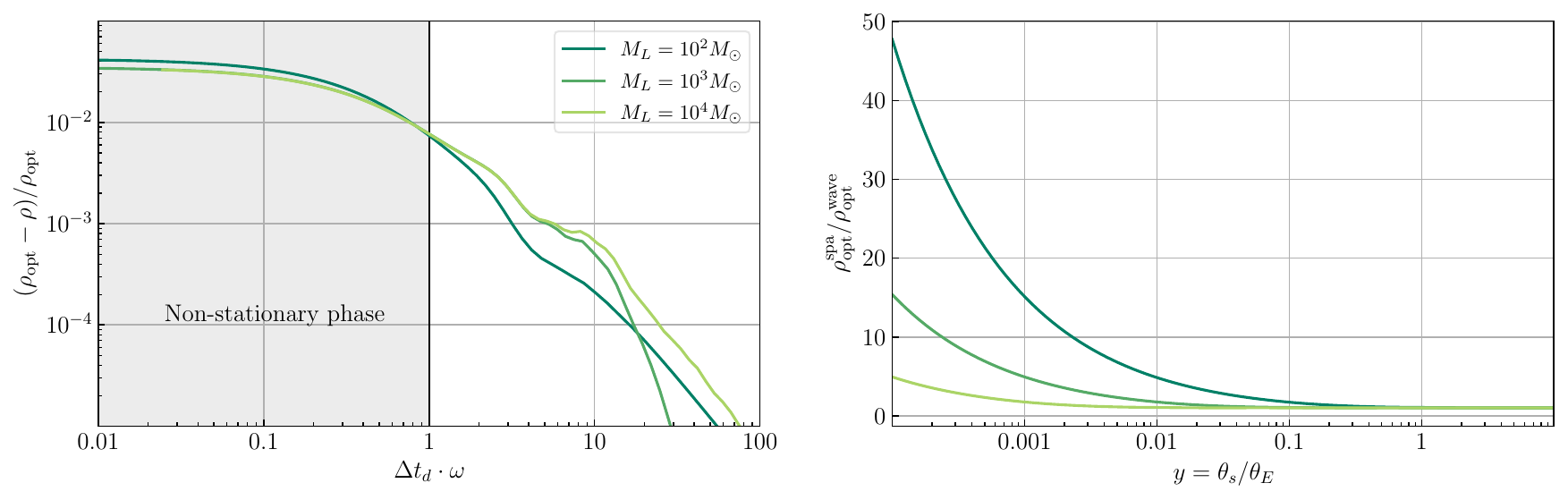}
 \caption{On the left we plot the relative difference between the optimal SNR of a lensed GW (solving the diffraction integral) and the matched filter SNR using a stationary phase approximation (SPA) lensed waveform as a template. We fix the GW orbital frequency ($\omega/2\pi=257$Hz) and vary the source-lens alignment ($y\subset[10^{-4},10]$) to get different time delays between the type I and II images $\Delta t_d$. We present the results for different lens masses satisfying the geometric optics approximation, $t_M\cdot\omega>1$. 
On the right, we compare optimal SNR computed using the SPA for the signal with the optimal SNR for the true wave optics signal as a function of $y$. 
The rest of the binary and lens parameters correspond to $q=1$, $z_S=0.5$, $z_L=0.1$, $\theta=0.3$, $\phi=0.4$, $\psi=1.5$, $\iota=\pi/3$ and $\chi_\text{eff}=\chi_p=e=0$.}
 \label{fig:snr_diff_spa}
\end{figure*}

To highlight this problem,
 we take  the limit that the source is nearly aligned with the a point mass lens so $\Delta t_d \rightarrow 0$.  To see the waveform distortion clearly, we
divide the Gaussian pulse of Fig. \ref{fig:shift_cosine} in two so as to create a sharp front
in the signal
\begin{align} \label{eq:half_gaussian}
  S(t)=\Theta(t)e^{-\frac{1}{2\sigma^2}\left(\frac{t}{t_M}\right)^2},
\end{align}
where $\Theta(t)$ is the Heaviside step function.
In this case, the lensed signal for the point mass lens shows an acausal tail at $\Delta T<0$ which can be traced back to the contribution of the type II image. This is in fact caused by
a breakdown in the stationary phase approximation performed to obtain Eq.~(\ref{Fgeom}).   
Mathematically, the breakdown occurs since the saddle point has 
principal directions that extend to both smaller and larger time delays. In the stationary phase approximation, these delays are extrapolated to infinity using a quadratic expansion.  In reality, the direction to smaller delays is bounded by the global minimum that corresponds to the type I image.  

Thus the full amplification factor $F$ never exhibits this false
superluminality and the first arrival of the front is always
given by the type I image.  While we have illustrated this for the the extreme case where $y\rightarrow 0$ in Fig.~\ref{fig:acausal}, the breakdown applies in principle  to any part, however small, of the type II signal
that appears in advance of the type I signal. 
This means that  lensing can distort the waveform beyond the stationary phase approximation even if $t_M \cdot \omega \gg 1$.

We now quantify whether this waveform distortion  will jeopardize matched filtering searches for binary sources in practice.  Since these signals are nearly monochromatic and in particular do not have a well defined temporal front to distort, the impact is in practice  negligible.

On the left panel of Fig. \ref{fig:snr_diff_spa} we plot the relative SNR difference as a function  of $\Delta t_d \cdot \omega$, which controls the  validity of the stationary phase limit equation (\ref{Fgeom}). We use  the exact expression of the amplification factor for a point lens to compute the lensed waveform (see e.g. \cite{Ezquiaga:2020spg}). The template in this example corresponds to the formula (\ref{Fgeom}). Each of the lines in the plot represents different lens masses, which have been chosen so that each examples satisfies  
$t_M\cdot\omega>1$. The lens mass together with the source position $y=\theta_s/\theta_E$ determine the time delay between the images $\Delta t_d$. Here we fix the frequency of the GW with the frequency at the inner most stable circular orbit (ISCO) for a $30-30M_\odot$ binary. We vary $y$ in order to change the time delay between the images, restrict to the range $y\subset[10
^{-4},10]$.
We see that the SNR degrades as $\Delta t_d \omega$ approaches unity from above but then this degradation rapidly saturates to a maximal value that is still relatively small and only weakly dependent on parameters.   Even as the source becomes perfectly aligned, the degradation does not become large since the superposition of the stationary phase approximation templates is not sensitive to delays that are shorter than the period of the wave.

\begin{figure}[t!]
\centering
\includegraphics[width = 0.95\columnwidth]{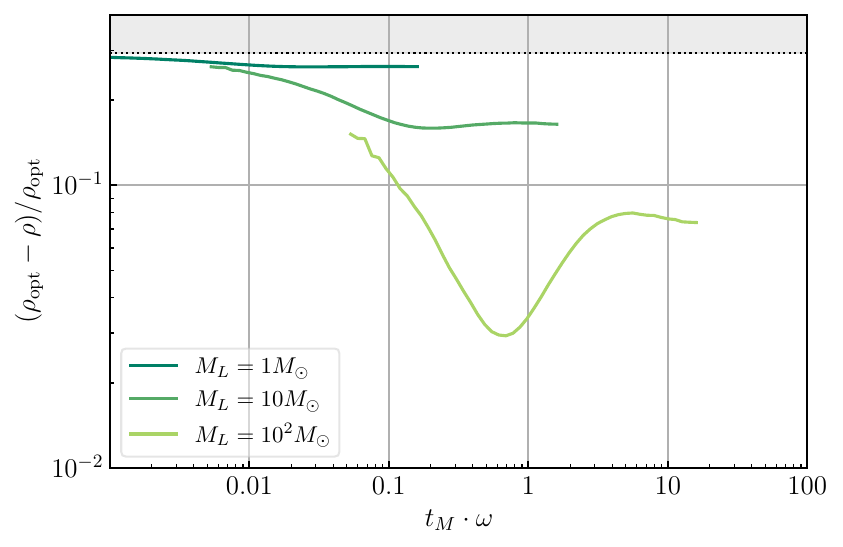}
 \caption{Relative difference between the optimal SNR of a lensed GW (solving the diffraction integral) and the matched filter SNR using a stationary phase approximation lensed waveform as a template. 
We fix the source-lens alignment in the strong lensing regime ($y=0.01$) and vary $\omega$ choosing different chirp masses ($\mathcal{M}_c\subset[2,600]M_\odot$). 
The rest of the parameters are the same as in Fig.~\ref{fig:snr_diff_spa}. The shaded region shows the limit in the SNR difference when $t_M\cdot\omega\to0$.}
 \label{fig:snr_diff_wave_geom}
\end{figure}

On the other hand, the stationary phase approximation for the magnification degrades without bound as $\Delta t_d \cdot \omega \rightarrow 0$.
While the SNR is independent of this mismatch of the amplitude of this template, if we instead falsely assume that the signal was magnified according to the stationary phase formula (\ref{Fgeom}) we would overestimate the SNR.  We present on the right panel of Fig. \ref{fig:snr_diff_spa} the SNR ratio of the optimal SNR computed using the stationary phase approximation 
against the one computed with the exact wave-optics formula. We present this quantity as a function of the true position of the source $y$. 
As the source-lens are more aligned ($y\to0$) the GW crosses closer to the caustic. Near the caustic the stationary phase limits breaks down and the magnification diverge as $\sqrt{\mu}\sim 1/y$. For example, for the binary of Fig.~\ref{fig:snr_diff_wave_geom} and a lens $M_L=100M_\odot$ at $y=10^{-4}$ the optimal SNR estimated with the stationary phase approximation will be 50 times larger than the wave-optics results. 
This amplification overestimation has implications for computing the probabilities of highly magnified GW events. 

More significant waveform distortions from the stationary phase approximation templates can occur in the limit where the wave period is smaller than the Schwarzschild crossing time, i.e. $t_M \cdot \omega < 1$ (recall that in strong lensing this only occurs in the regime violating the stationary phase approx. $\Delta t_d\cdot\omega<1$). 
In Fig. \ref{fig:snr_diff_wave_geom},
we show  the relative SNR difference as a function of $t_M\cdot\omega$, fixing the source position in the strong lensing regime ($y=0.01$) and varying the frequency by changing the chirp mass ($\mathcal{M}_c\subset[2,600]M_\odot$).  In this plot it is clear that independently of the lens mass the SNR difference tends to a maximum saturation limit (dotted line). 
This maximum SNR difference is determined by the wave optics limit $t_M\cdot \omega \rightarrow 0$ in which there is no phase shift once all paths are superimposed. Note that in this limit the amplification factor, cf. Eq.~(\ref{Feq}), $F\to1$ and the effect of lensing is negligible. 
On the contrary, the stationary phase approximation will predict a phase shift of $\pi/4$, just because when $\Delta t_d\to0$ we are adding two images with phase 0 and $\pi/2$. One can check that this saturation limit is nothing but $1-\cos(\pi/4)$ and should be universal for detected signals dominated by the inspiral phase. 
Since we are restricting the analysis to current ground-based detectors (O3 sensitivity specifically), small frequencies ($\lesssim10$Hz) are not detected and this is why the lines cut on the left end. The non-monotonic behavior of the SNR around $t_M \cdot \omega \sim 1$  (most notably seen in $M_L=10^3M_\odot$) is caused by a similar behavior of the wave-optics phase (see Eq.~(5) of \cite{Ezquiaga:2020spg} for the specific formula), but in all the cases the phase vanishes when $\omega\to0$ as shown in Fig.~2 of \cite{Ezquiaga:2020spg}.

\section{Antenna Response and Multiple Detectors}
\label{app:antenna}

In the main text, we considered degeneracies between lensing phase shifts and polarization angle 
$\psi$ for a single detector. In this Appendix we first clarify conventions for specifying $\psi$ and then proceed to the case of
multiple detectors.

\subsection{Antenna pattern}

First we consider a single detector whose orientation with respect to the source is approximately fixed, at least during the duration of the signal. 
There are several conventions for defining the antenna pattern of the response of such a detector to a gravitational 
wave.   These conventions  differ by the choice of angles that characterize the gravitational wave polarization state axes relative to the detector arms.   In this Appendix we clarify the relationship between these conventions.  

The radiation polarization is
defined in the so-called radiation frame by the right-handed triad 
$\{ \vec{x}_r, \vec{y}_r, -\vec{n} \}$.  Recall that $-\vec{n}$ is the propagation direction since $\vec{n}$ is the direction from detector to source.  In our case, depicted in Fig.~\ref{fig:angles}, $\vec{x}_r$ is defined by the projection of $\vec{J}$ on the plane transverse to $-\vec{n}$ 
(or equivalently $\vec{n}$) but we leave
the notation general here.
The polarization states are therefore defined through the polarization tensors
\begin{eqnarray}
{\bf e}_+&=& \vec{x}_r \otimes \vec{x}_r - \vec{y}_r \otimes \vec{y}_r,\nonumber\\
{\bf e}_\times &=&  \vec{x}_r \otimes \vec{y}_r + \vec{y}_r \otimes \vec{x}_r.
\end{eqnarray}
The transverse traceless metric tensor describing gravitational waves is then given by $h_{ij}=h_{+}{\bf e}_{+,ij}+ h_{\times} {\bf e}_{\times, ij}$, with $h_+$ and $h_\times$ the polarization amplitudes in the radiation frame.

The detector is described by the arms of the interferometer which we take here to be orthogonal, $\vec{x}_d$, $\vec{y}_d$ forming the right-handed triad $\{ \vec{x}_d, \vec{y}_d, \vec{z}_d \}$ and measures $``+"$ detector-frame polarization described by 
\begin{eqnarray}
\bf{D} &=& \frac{1}{2} ( \vec{x}_d \otimes \vec{x_d} - \vec{y}_d \otimes \vec{y_d}).
\end{eqnarray}
In general, the response of the detector to the radiation frame polarization states is given by the antenna response
\begin{eqnarray}
F_+ &=& {\rm Tr}[ {\bf D} \, {\bf  e}_+ ],\nonumber\\
F_\times &=& {\rm Tr} [{\bf D}\,  {\bf  e}_\times].
\end{eqnarray}
These response functions therefore depend on the relative alignments of the radiation and detector frames.   Note that since the two coordinate systems are not aligned before rotation, the matrix operations are in 3 dimensions.  

The relative alignments themselves can be described by Euler angles that rotate the detector frame onto the
radiation frame.   The angles are uniquely determined once a convention for the Euler angles  and the alignment of any two axes are chosen, since the third is specified by the right-handedness of both coordinate systems.   
We choose the $zyz$ Euler rotation angles $\{\phi,\theta,\psi\}$, with the rotation matrix $R(\phi,\theta,\psi)$, where each angle represents a counterclockwise rotation around the currently specified axis after the previous rotations.   
We then fix the convention by
determining which detector axis aligns with $\vec{n}$
and $\vec{x}_r$ after rotation. Next, we mention three conventions typically found in the literature.

\subsubsection{$\{ \vec{z}_d \rightarrow \vec{n}, \vec{x}_d 
\rightarrow \vec{x}_r \}$ Convention}

The conventions of this paper are to describe the relative orientation of the radiation and detector frames by determining the $zyz$ Euler angles that rotate  
$\vec{z}_d \rightarrow \vec{n}$, $\vec{x}_d 
\rightarrow \vec{x}_r$.   Note that $\vec{y}_d \rightarrow -\vec{y}_r$ as a consequence of the right-handedness of both systems.  
This convention has the advantage that
the first two rotations $\phi$ and $\theta$, which rotate $\vec{z}_d$ to $\vec{n}$ are the
coordinates of the source position with
polar angle $\theta$ and azimuthal angle $\phi$.  The final rotation by $\psi$ is counterclockwise around $\vec{n}$ and aligns
the {\it rotated} $\vec{x}_d \rightarrow \vec{e}_{\theta}$ to $\vec{x}_r$.   It is important to note that
this is not the same as the angle defining the separation between the {\it projection} of $\vec{x}_d$ onto the plane transverse to $\vec{n}$ and $\vec{x}_r$ due to the first
two rotations.  
In this convention, the radiation frame is then obtained as:
\begin{eqnarray}
\vec{x}_r &=& {\bf R}(\phi,\theta,\psi) (\vec{x}_d),
\nonumber\\
\vec{y}_r &=&  {\bf R}(\phi,\theta,\psi)  (-\vec{y}_d),
\nonumber\\
- \vec{n} &=&  {\bf R}(\phi,\theta,\psi)  (-\vec{z}_d).
\end{eqnarray}
The antenna pattern in terms of these rotation angles is then given by:
\begin{align}
F_+ ={} & \frac{1}{2} (1+ \cos^2\theta) \cos(2 \phi) \cos(2 \psi) \nonumber\\{} &
- \cos\theta \sin(2\phi) \sin(2\psi), \nonumber\\
F_\times ={} & \frac{1}{2} (1+ \cos^2\theta) \cos(2 \phi) \sin(2 \psi)
\nonumber\\{} &
+ \cos\theta \sin(2\phi) \cos(2\psi) .
\end{align}
For example if the source is overhead the detector
so that $\theta=0,\phi=0$, then $\psi=0$ means that the $\vec{x}_r=\vec{x}_d$ and $\vec{y}_r=-\vec{y}_d$ without further rotation so that $F_+=1$
and $F_\times =0$. This is the same antenna pattern functions obtained in \cite{Sathyaprakash:2009xs}.

\subsubsection{$\{ \vec{z}_d \rightarrow \vec{n}, \vec{y}_d \rightarrow x_r \}$ Convention }

Another option, chosen by LIGO \cite{Anderson:2000yy,ligo_antenna}, is to define 
the angles that rotate 
\begin{eqnarray}
\vec{x}_r &=& {\bf R}(\phi,\theta,\psi_{\rm L}) (-\vec{y}_d),
\nonumber\\
\vec{y}_r &=& {\bf R}(\phi,\theta,\psi_{\rm L}) (-\vec{x}_d),
\nonumber\\
-\vec{n} &=& {\bf R}(\phi,\theta,\psi_{\rm L} ) (-\vec{z}_d).
\end{eqnarray}
The first two rotations are the same as our convention, but the final rotation differs, so that the response is now
\begin{align}
F_+ ={}& -\frac{1}{2} (1+ \cos^2\theta) \cos(2 \phi) \cos(2 \psi_{\rm L}) 
\nonumber\\{} &
- \cos\theta \sin(2\phi) \sin(2\psi_{\rm L}), \nonumber\\
F_\times ={}& \frac{1}{2} (1+ \cos^2\theta) \cos(2 \phi) \sin(2 \psi_{\rm L})
\nonumber\\{} &
- \cos\theta \sin(2\phi) \cos(2\psi_{\rm L}) .
\end{align}
 Note that, for the same choice of detector and radiation frames, the response of the detector does not depend on how we choose to represent the rotations so that $\psi_{\rm L}= -\psi + 3\pi/2$.
 This accounts for the rotation required to align $-\vec{y}_d$ instead of 
$\vec{x}_d$ with $\vec{x}_r$.  Correspondingly
for an overhead source $\theta=0$, $\phi=0$, then
$\psi_{\rm L}=0$ means that $\vec{x}_r=-\vec{y}_d$ and $\vec{y}_r=-\vec{x}_d$ so $F_+=-1$
and $F_\times =0$.

Note that this convention also arises  if the
Euler rotations are defined to rotate 
$\vec{z}_d \rightarrow -\vec{n}$ and $\vec{x}_d
\rightarrow \vec{x}_r$ using a $zxz$ Euler rotation, followed by a relabeling 
of the first two Euler angles into sky coordinates keeping the final angle as $\psi_{\rm L}$.
In this interpretation the reversal of sign for
$\psi_{\rm L}$ is related to the $\vec{n}\rightarrow -\vec{n}$ inversion and the $x-y$ inversion comes
from the relationship between $zxz$ Euler angles $\{\alpha',\beta',\gamma' \}$ and $zyz$ angles $\{ \alpha,\beta,\gamma \}$:  $\alpha=\alpha'-\pi/2$, $\beta=\beta'$, $\gamma=\gamma'+\pi/2$.   The 
composition of the $\vec{n}$ inversion and the Euler
rotation flips $x$ and $y$ for a source located at
$\theta=0,\psi=0$ since $\beta=\pi$.
Since a orientation convention must specify {\it both} the Euler convention and the alignment after rotation, we represent this convention in the $zyz$
scheme here.

\subsubsection{$\{ \vec{z}_d \rightarrow -\vec{n}, \vec{x}_d \rightarrow \vec{x}_r \}$ Convention}

Given that $-\vec{n}$ is the
propagation direction, we can also choose to align it to $\vec{z}_d$ and 
define
\begin{eqnarray}
\vec{x}_r &=& {\bf R}(\Phi,\Theta,\psi_{\rm R}) (\vec{x}_d),
\nonumber\\
\vec{y}_r &=& {\bf R}(\Phi,\Theta,\psi_{\rm R}) (\vec{y_d}),
\nonumber\\
-\vec{n} &=& {\bf R}(\Phi,\Theta,\psi_{\rm R} ) (\vec{z}_d).
\end{eqnarray}
Notice that the first two rotations no longer correspond to the source sky position but are
related by $\Theta=\pi -\theta$ and $\Phi=\phi+\pi$.
In this case 
\begin{align}
F_+ ={}& \frac{1}{2} (1+ \cos^2\Theta) \cos(2 \Phi) \cos(2 \psi_{\rm R}) 
\nonumber\\{} &
- \cos\Theta \sin(2\Phi) \sin(2\psi_{\rm R}) \nonumber\\
 ={}& \frac{1}{2} (1+ \cos^2\theta) \cos(2 \phi) \cos(2 \psi_{\rm rad}) 
 \nonumber\\{} &
 + \cos\theta \sin(2\phi) \sin(2\psi_{\rm R}), \nonumber\\
F_\times ={}& -\frac{1}{2} (1+ \cos^2\Theta) \cos(2 \Phi) \sin(2 \psi_{\rm R}) 
\nonumber\\{} &
- \cos\Theta \sin(2\Phi) \cos(2\psi_{\rm R})\nonumber\\ 
={}&
-\frac{1}{2} (1+ \cos^2\theta) \cos(2 \phi) \sin(2 \psi_{\rm rad}) 
\nonumber\\{} &
+ \cos\theta \sin(2\phi) \cos(2\psi_{\rm R}) ,
\end{align}
and hence $\psi_{\rm R}=-\psi$ which reflects the
coordinate inversion so that $\psi$ is a clockwise rotation around $-\vec{n}$.
In this convention, for an overhead source
$\theta=0$, $\phi=0$, if $\psi_R=0$ then $\vec{x}_r=\vec{x}_d$ and $\vec{y}_r=-\vec{y}_d$ and $F_+=1$, $F_\times=0$ as in our convention but finite values of the $\psi$ rotation would differ due to their  opposite sign. Note that in some references, e.g.~\cite{Maggiore:1900zz}, the antenna pattern functions are expressed in terms of the angles $\{\Theta,\Phi\}$ of the propagation direction $-\vec{n}$. 

\subsection{Multiple Detectors}

Now consider the case of multiple detectors, e.g.~on the surface of the Earth, having different relative orientations among them. We can relate their antenna responses with respect to any other reference frame. This is useful when computing the source localization with respect to the Earth fixed frame, a fixed coordinate system at the center of the Earth with $z$-axis pointing to the North pole, $x$-axis pointing from the origin to the intersection of the equator
and prime meridian and $y$-axis completing the right-handed coordinate system \cite{ligo_earth_frame}. It is to be noted that the Earth fixed frame follows the rotation of the Earth.

Let us now describe the  orientation of the $n$th detector relative to  a reference coordinate system, e.g. the Earth fixed frame now considered to be the $\{\vec{x}_d,\vec{y}_d,\vec{z}_d \}$ frame of the previous section, by $zyz$ Euler angles
\begin{align}
    \vec{x}_{dn} & = R(\alpha_n,\beta_n,\gamma_n) \vec{x}_d, \nonumber\\
        \vec{y}_{dn} & = R(\alpha_n,\beta_n,\gamma_n)
        \vec{y}_d.
\end{align}
Their antenna responses are then given by
\begin{eqnarray}
F_{+n} &=& {\rm Tr}[ {\bf D}_n \, {\bf  e}_+ ],\nonumber\\
F_{\times n} &=& {\rm Tr} [{\bf D}_n\,  {\bf  e}_\times],
\end{eqnarray}
where
\begin{eqnarray}
{\bf D}_n &=& \frac{1}{2} ( \vec{x}_{dn} \otimes \vec{x}_{dn} - \vec{y}_{dn} \otimes \vec{y}_{dn}).
\end{eqnarray}
These responses then define the phase seen by each detector for a GW multipole $(\ell,m)$ in a straightforward generalization of Eq.~(\ref{AvsAngles}):
\begin{equation}
    \chi_{\ell m}^n = {\rm arctan}[F_{+n}, f_{\ell m} F_{\times n}],
\end{equation}
which now depend on the 3 angles of the reference frame $\{\theta,\phi,\psi \}$ and the 3 angles of each detector $\{\alpha_n,\beta_n,\gamma_n \}$.

It is now straightforward to ask the question whether a single global change to $\psi$ by $\Delta \psi_{22}$ preserves the $\pi/2$ lensing phase shift independently of the relative orientation of the detectors. 
For $\iota=0,\pi$ where the emission is circularly polarized, this is always the case and the degeneracy remains as exact as with a single detector. 

\begin{figure}[t!]
\centering
\includegraphics[width = 0.95\columnwidth]{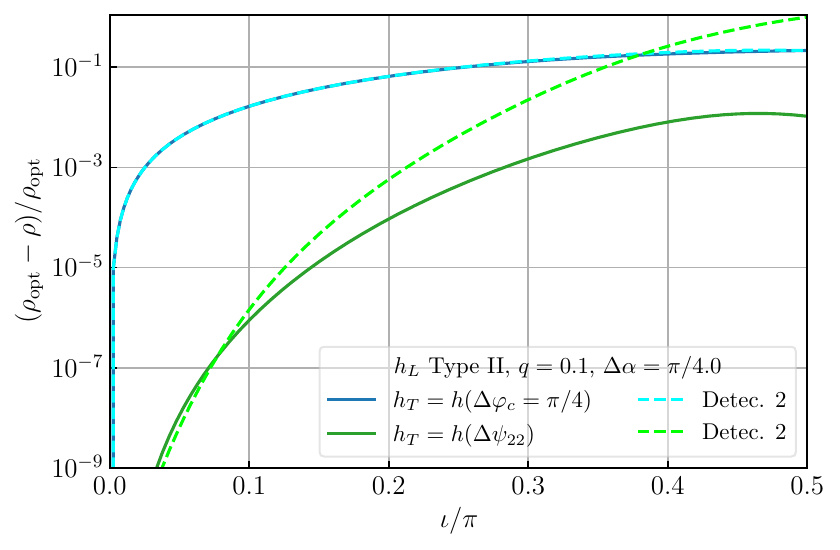}
 \caption{Relative difference between the optimal SNR of a lensed GW and the matched filter SNR using templates with a shift in the coalescence phase $\Delta\varphi_c$ or a shift in the orientation angle $\Delta\psi_{22}$ w.r.t. a fixed detector, ``detector 1".  
Dashed lines represent the SNR seen by a ``detector 2" rotated by $\Delta\alpha=\pi/4$ with respect to the fix detector 1 of the solid lines. We plot the SNR difference as a function of the inclination $\iota$ for a non-spinning, circular binary with mass ratio $q=1/10$.  
The rest of the parameters are the same as in Fig. \ref{fig:snr_diff}.}
 \label{fig:snr_diff_multiple_detectors}
\end{figure}

The maximal discrepancy is
the edge on case $\iota \rightarrow \pi/2$ where the emission is linearly polarized and multiple detectors can 
distinguish between rotation and a phase shift.   For example if a pair of detectors have a relative orientation
$\alpha_1=0$, $\alpha_2=\pi/4$
with $\beta_{1,2}=\gamma_{1,2}=0$, we have the maximal discrepancy where  the phases will differ by $\pi/2$ so that no global change in $\psi$ can introduce a type II phase shift in both detectors. 
We can see this explicitly in Fig. \ref{fig:snr_diff_multiple_detectors}, where we present the SNR difference between such two detectors as a function of the inclination. 
We have verified that for any other combination of $\alpha_{1,2}$, $\beta_{1,2}$ and $\gamma_{1,2}$ the maximal SNR difference for a $q=0.1$ binary with $\iota\leq\pi/3$ and $\iota\leq0.4\pi$ is less than $\sim7\%$ and $\sim30\%$ respectively.

\bibliography{RefModifiedGravity}

 \end{document}